\documentclass{emulateapj}
\usepackage[bottom]{footmisc}
\usepackage{color,amsmath,enumitem}
\usepackage{multirow}

\shorttitle{Bayesian Analysis of Multiple Populations I: Statistical and Computational Methods}
\shortauthors{Stenning et al.}

\begin{document}

\newcommand{\markup}[1]{\textbf{\textcolor{red}{{\fontfamily{garamond}\selectfont{#1}}}}}
\renewcommand*{\thefootnote}{\fnsymbol{footnote}}

\newcommand{\MSun}{$M_{\text{Sun}}$}
\newcommand{\bmu}{\mbox{\boldmath{$\mu$}}}
\newcommand{\balpha}{\mbox{\boldmath{$\alpha$}}}
\newcommand{\bSigma}{\mbox{\boldmath{$\Sigma$}}}
\newcommand{\bTheta}{\mbox{\boldmath{$\Theta$}}}
\newcommand{\bOmega}{\mbox{\boldmath{$\Omega$}}}
\newcommand{\btheta}{\mbox{\boldmath{$\theta$}}}
\newcommand{\bphi}{\mbox{\boldmath{$\phi$}}}
\newcommand{\bPhi}{\mbox{\boldmath{$\Phi$}}}
\newcommand{\bM}{\mbox{\boldmath{$M$}}}
\newcommand{\bR}{\mbox{\boldmath{$R$}}}
\newcommand{\bZ}{\mbox{\boldmath{$Z$}}}
\newcommand{\bX}{\mbox{\boldmath{$X$}}}
\newcommand{\bG}{\mbox{\boldmath{$G$}}}
\newcommand{\bF}{\mbox{\boldmath{$F$}}}
\newcommand{\bV}{\mbox{\boldmath{$V$}}}
\newcommand{\bGMS}{\mbox{\boldmath{$G$}}_{\textrm{MS/RG}}}
\newcommand{\bGWD}{\mbox{\boldmath{$G$}}_{\textrm{WD}}}
\newcommand{\bY}{\mbox{\boldmath{$\Psi$}}}
\newcommand{\diag}{\textrm{diag}}
\newcommand{\thetaage}{\mbox{$\theta_{\text{age}}$}}
\newcommand{\thetaFeH}{\mbox{$\theta_{\text{[Fe/H]}}$}}
\newcommand{\thetamod}{\mbox{$\theta_{m - M_V}$}}
\newcommand{\thetaabs}{\mbox{$\theta_{A_V}$}}
\newcommand{\thetaY}{\mbox{$\theta_{Y}$}}
\newcommand{\thetaC}{\mbox{$\theta_{C}$}}
\newcommand{\phiY}{\mbox{$\phi_{Y}$}}
\newcommand{\phiprog}{\mbox{$\phi_{\text{prog\,age}}$}}
\newcommand{\phiTeff}{\mbox{$\phi_{T_{\text{eff}}}$}}
\newcommand{\phiradius}{\mbox{$\phi_\text{radius}$}}
\newcommand{\philogg}{\mbox{$\phi_{\log g}$}}
\newcommand{\bxi}{\mbox{\boldmath{$\Xi$}}}
\newcommand{\bDelta}{\mbox{\boldmath{$\Delta$}}}
\newcommand{\indep}{\stackrel{indep}{\sim}}
\newcommand{\lambdak}{\mbox{$\lambda_{k}$}}
\newcommand{\changes}[1]{{#1}}

\title{Bayesian Analysis of Two Stellar Populations in Galactic Globular Clusters I: Statistical and Computational Methods}
\author{D.C. Stenning*$^{1}$, R. Wagner-Kaiser$^{2}$, E. Robinson$^{3}$, D.A. van Dyk$^{4}$, T. von Hippel$^{5}$, A. Sarajedini$^{2}$, N. Stein$^{6}$}
\altaffiltext{1}{Sorbonne Universit\'{e}s, UPMC-CNRS, UMR 7095, Institut d'Astrophysique de Paris, F-75014 Paris, France}
\altaffiltext{2}{Bryant Space Center, University of Florida, Gainesville, FL}
\altaffiltext{3}{Argiope Technical Solutions, Florida, USA}
\altaffiltext{4}{Imperial College London, London, UK}
\altaffiltext{5}{Center for Space and Atmospheric Research, Embry-Riddle Aeronautical University, Daytona Beach, FL}
\altaffiltext{6}{The Wharton School, University of Pennsylvania, Philadelphia, PA}
\email{*stenning@iap.fr}


\begin{abstract}
We develop a Bayesian model for globular clusters composed of multiple stellar populations, extending earlier statistical models for open clusters composed of simple (single) stellar populations \citep[e.g.,][]{vanDyk_2009, Stein_2013}.  Specifically, we model globular clusters with two populations that differ in helium abundance.  Our model assumes a hierarchical structuring of the parameters in which physical properties---age, metallicity, helium abundance, distance, absorption, and initial mass---are common to (i) the cluster as a whole or to (ii) individual populations within a cluster, or are unique to (iii) individual stars.  An adaptive Markov chain Monte Carlo (MCMC) algorithm is devised for model fitting that greatly improves convergence relative to its precursor non-adaptive MCMC algorithm.  Our model and computational tools are incorporated into an open-source software suite known as BASE-9. We use numerical studies to demonstrate that our method can recover parameters of two-population clusters, and also show model misspecification can \changes{potentially be identified}.  As a proof of concept, we analyze the two stellar populations of globular cluster NGC 5272 using our model and methods. \changes{(BASE-9 is available from GitHub: \url{https://github.com/argiopetech/base/releases}).}
\end{abstract}


\section{Introduction}\label{Intro}

Globular clusters have long been used as probes of the formation and evolution of galaxies \citep[e.g.,][]{Sandage_1962, Searle_1978, Janes_1983, Lee_2001, Marin-Franch_2009, Forbes_2010}.  Past work on globular clusters has largely assumed that they consist of simple stellar populations, i.e., single stellar populations.  However, within the past decade, this assumption has come under scrutiny as numerous studies have produced evidence that globular clusters in fact host multiple distinct stellar populations \citep[e.g.,][]{Bedin_2004, Gratton_2004, Carretta_2006, Villanova_2007, Piotto_2007, Piotto_2009, Milone_2012}.  The implication is that most globular clusters have undergone multiple epochs of star formation \citep{Piotto_2015}.  As a result, globular clusters should be viewed as a mixture of two or more simple stellar populations.

When working with photometric magnitudes, the multiple populations are most prominent in ultraviolet (UV) color-magnitude diagrams (CMDs).  While previous studies focused on visual wavelengths, recent high-quality UV photometric data from the Hubble Space Telescope (HST) allow us to better investigate the presence of multiple stellar populations.  In fact, the vast majority of globular clusters that have been studied in the UV to sufficient accuracy display characteristics that can be attributed to multiple populations \citep{Piotto_2015}.  

Despite the substantial resources devoted to observing globular clusters and developing stellar evolution models, the methods used to fit costly models to expensive data typically neither take advantage of modern statistical methods nor incorporate astrophysical knowledge.  Investigators often use a ``chi-by-eye'' approach of plotting stellar evolution models on top of observed data and adjusting the parameters with the aim of achieving an acceptable fit, where the goodness-of-fit is determined by visual inspection.  Such approaches yield inaccurate results and cannot capture uncertainties in the model fits even when analyzing single-population star clusters \citep{vanDyk_2009, Jeffery_2015}.  At best, visual model fits are inherently subjective and difficult to reproduce, and rely on two-dimensional projections of the data.  When studying globular clusters that host multiple stellar populations, ``chi-by-eye'' fails completely because the populations may exhibit only small differences in a few parameters, and the stellar populations cannot be cleanly separated in the plotted CMDs.  

\begin{figure*}[t!]
\begin{center}
\includegraphics[width=0.8\textwidth]{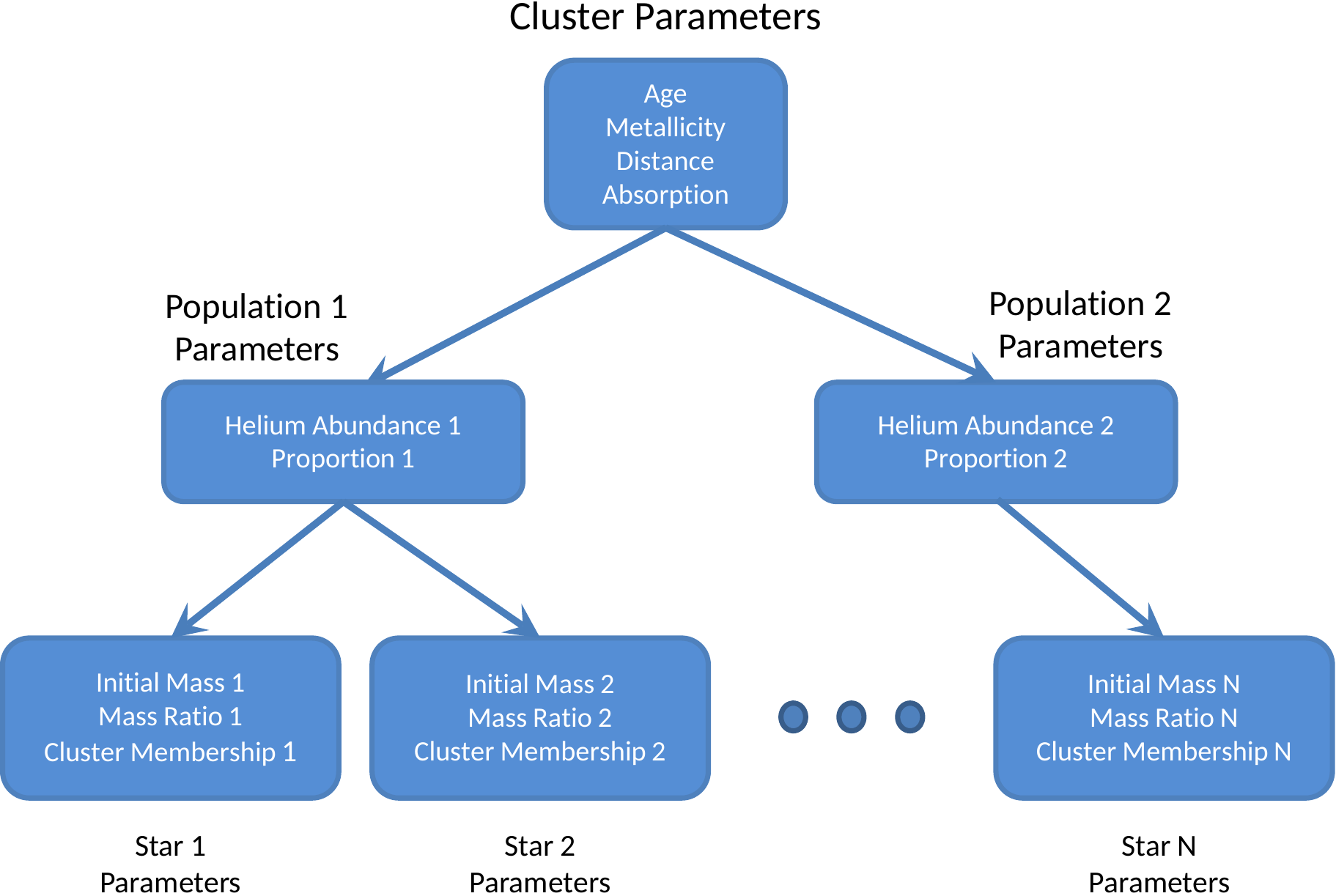}
\end{center}
\caption[Hierarchy of cluster, population, and stellar parameters for a two-population globular cluster] {Hierarchy of cluster, population, and stellar parameters for a two-population globular cluster.  The cluster parameters---age, metallicity, distance, and absorption---are common to all stars in the cluster.  The population parameters---helium abundance and the proportion of stars in a particular population---are common to all stars in a population but may be different between populations.  The stellar parameters---initial mass, mass ratio, and cluster membership indicator---are allowed to vary on a star-by-star basis.}
\label{fig:MultipopHierarchy}
\end{figure*}

In this article we present a Bayesian model for globular clusters that harbor two stellar populations, hereafter ``two-population globular clusters.''  This model is an extension of the model for simple stellar populations developed by \citet{vonHippel_2006}, \citet{DeGennaro_2009}, \citet{vanDyk_2009}, and \citet{Stein_2013}.  Our two-population model assumes that a globular cluster hosts two stellar populations that differ only in helium abundance.  This results in a hierarchy of properties with parameters associated either to individual stars, stellar populations, or the globular cluster as a whole. This hierarchy is illustrated in Figure~\ref{fig:MultipopHierarchy}, and the parameters are defined in Table~\ref{tbl:multipop_params}; the notation and terminology in Table~\ref{tbl:multipop_params} is introduced in Section~\ref{subsec:Notation}.  

Our statistical model accounts for measurement errors, field star contamination, and the possibility of stellar binaries.  Adopting a Bayesian approach for model fitting provides principled and reproducible estimates and uncertainties on all parameters.  Future work will incorporate variations between the light element abundances and other population-level characteristics, but for this first study we choose to limit our attention to a single parameter that varies between the populations and is expected to significantly alter the morphology of the CMD.  Estimating the difference in helium abundance provides insight into the possible mechanisms that produce multiple-population clusters.  To fit our two-population model, we implement an adaptive Metropolis algorithm \citep[e.g.,][]{Haario_2001, Roberts_2009, Rosenthal_2011}.  This algorithm has the benefit of improving convergence compared to a standard (non-adaptive) Metropolis algorithm, without requiring significant tuning by the user.  

\begin{table*}[t!]
\begin{center}
\caption{Two-Population Model Parameters}
\label{tbl:multipop_params}
\begin{tabular}{llr}
\hline
 {\small Parameter} & {\small Description} & {\small Notation} \\  \hline \\
	{\bf Cluster Parameters} \\ \\
	{\it Age} & $\log_{10}$ of cluster age in years & \thetaage  \\ 
	{\it Distance} & distance modulus in mag  & \thetamod  \\
	{\it Absorption} & absorption in the V-band in mag & \thetaabs  \\ 
	{\it Metallicity}  & $\log_{10}$ of iron-to-hydrogen ratio relative to Sun in dex \qquad \qquad & \thetaFeH \vspace{0.1cm} \\ \hline \\
	{\bf Population Parameters} \qquad \qquad \\ \\
	{\it Proportion} &  proportion of stars from a population & $\phi_{pk}$ \\
	{\it Helium Abundance} & mass fraction of helium & $\phi_{Yk}$ \vspace{0.1cm} \\ \hline \\
	{\bf Stellar Parameters} \\ \\
 	{\it Initial Mass} &  Zero Age Main Sequence mass in solar units, $M_{\odot}$ & $M_i$\\
	{\it Mass Ratio} & ratio of secondary to primary initial masses & $R_i$ \\
	{\it Cluster Membership} & indicator for cluster membership & $Z_i$ \vspace{0.1cm} \\
\hline
\end{tabular}
\end{center}
\end{table*}

Our model and methods are incorporated into an open-source software suite known as BASE-9 for {\bf B}ayesian {\bf A}nalysis of {\bf S}tellar {\bf E}volution with {\bf 9} Parameters.  A combination of several computer-based stellar evolution models is used to predict a star's photometric magnitudes given a set of stellar evolution parameters: age, distance, absorption, metallicity, helium abundance, and initial mass.  To recover star cluster parameters from photometric data, BASE-9 includes sophisticated MCMC routines for model fitting.  BASE-9 is available as open source code from GitHub (\url{https://github.com/argiopetech/base/releases}), and is also available as executables through Amazon Web Services.  Additional technical details can be found in the BASE-9 Manual \citep{vonHippel_2014}.

For main sequence and red giant stars, BASE-9 gives users a choice of the state-of-the-art models by \citet[][and updated at \href{http://stellar.dartmouth.edu/~models/}{http://stellar.dartmouth.edu/$\sim$models/}]{Dotter_2008} and the commonly used models of \citet{Girardi_2000} and \citet{Yi_2001}.  Other models are available for white dwarfs, as well as for the initial-final mass relations that bridge the stages of stellar evolution.  These models are not pertinent to the current discussion because our analyses of two-population globular clusters are limited to main sequence through red giant branch stars.   

The rest of this article is divided into five sections.  In Section~\ref{sec:MuliPopModel} we present our statistical model for two-population globular clusters.  In Section~\ref{sec:MultipopStatComp} we discuss the computational challenges involved with fitting this model, and show how adaptive MCMC techniques improve convergence.  In Section~\ref{sec:MultipopNumRes} we illustrate the capabilities of our model and methods using a series of numerical studies.  In Section~\ref{sec:multipop_data_analysis} we present the results of fitting our two-population model to NGC 5272.  Finally, in Section~\ref{sec:MultiPopDiscussion} we summarize our results and discuss directions of future research.


\section{Statistical Model for Two-Population Globular Clusters}
\label{sec:MuliPopModel}

\subsection{Bayesian Modeling}

Bayesian methods offer a principled, probability-based approach for combining information from the current data and our prior knowledge.  They require a {\it likelihood function}---the distribution of the data given the model parameters.  The likelihood function is the primary statistical tool for assessing the viability of a parameter value vis-\`{a}-vis the observed data under a postulated statistical model.  The knowledge we have about the model parameters before considering the current data is specified in a {\it prior distribution}.  Past and current information are combined in the {\it posterior distribution} of the parameters, which is related to the likelihood function and the prior distribution through Bayes' theorem.  With generic data and model parameters represented by $Y$ and $\psi$, Bayes' theorem gives the posterior distribution as    
\begin{equation}
P(\psi | Y) = \frac{P(Y | \psi)P(\psi)}{P(Y)},
\end{equation}
where $P(Y | \psi) \equiv L(\psi | Y)$ is the likelihood function and $P(\psi)$ the prior distribution.  The term $P(Y)$, sometimes called the ``evidence,'' is a normalizing constant which makes $P(\psi | Y)$ a proper probability distribution.  The posterior distribution provides a summary of the combined information in the data and our prior knowledge and can be used to derive parameter estimates and uncertainties.  

To build a Bayesian model for two-population globular clusters, we start by defining necessary notation and terminology in Section~\ref{subsec:Notation}.  In Section~\ref{subsec:lik1} we construct a preliminary likelihood function for a simple stellar population that accounts for measurement error, the presence of field stars, and the possibility of binary star systems.  We extend this model to allow for two-stellar populations in Section~\ref{subsec:lik2}, and specify the full prior distribution in Section~\ref{subsec:priors}.

\subsection{Notation}
\label{subsec:Notation}

For each star in a dataset we obtain calibrated photometric magnitudes using at least two filters.  Following \citet{DeGennaro_2009}, \citet{vanDyk_2009} and \citet{Stein_2013}, we refer to the observed photometric magnitude in filter $j$ for star $i$ as $x_{ij}$  for $j=1,\dots, n$ and $i=1,\ldots, N$, where $N$ is the number of stars in the dataset and $n$ is the number of filters.  The observed photometric magnitudes for star $i$ are tabulated in the column vector $\bX_i = (x_{i1},\ldots,x_{in})^\top$, and the known (independent) Gaussian measurement errors in the (diagonal) variance-covariance matrix, $\bSigma_i$. 

As discussed in Section~\ref{Intro}, our statistical model is based on a hierarchy of parameters.  We refer to the parameters that are common to cluster stars---specifically age, metallicity, distance, and absorption---as {\it cluster parameters}.  These parameters are collected in the vector $\bTheta = (\thetaage,\thetaFeH,\thetamod,\thetaabs)$.  We refer to the parameters that are common to all stars belonging to a stellar population, but that vary from population to population within a cluster, as {\it population parameters}.  We assume that only helium abundance differs between the populations; helium abundance and the proportion of stars for population $k$, denoted $\phi_{Yk}$ and $\phi_{pk}$, respectively, are the population parameters.  (When discussing simple stellar populations we denote the single helium abundance with $\phi_{Y}$.)  We refer to the population with the lower helium abundance as ``Population 1," and that with the higher abundance as ``Population 2." (This should not be confused with the traditional use of Population I versus Population II stars.)  As a result, assuming that the two stellar populations result from two epochs of star formation, Population 1 corresponds to the first generation of stars and Population 2 to the second generation.  For now the only {\it stellar parameter} specific to star $i$ is its initial mass, $M_i$.  (Two more are specified below.)  The computer-based stellar evolution model, $\bG$, takes $(M_i, \bTheta, \phiY)$ and outputs a  $1 \times n$ vector of predicted photometric magnitudes for a star with those parameters.  We express the vector of predicted magnitudes as  $\bG(M_i, \bTheta, \phiY)$.  For this study, $\bG$ are the updated \citet{Dotter_2008} models that include HST UV magnitudes.
 
\subsection{Simple Stellar Populations}
\label{subsec:lik1}

Before considering the likelihood function for a two-population globular cluster, we first consider a ``preliminary'' likelihood function for a simple stellar population.  Following \citet{vanDyk_2009}, we account for unresolved binaries because the added luminosity of a binary companion shifts a star off the main sequence on the CMD, which can result in systematic errors if not properly handled.  We thus treat every observed star as a possible binary system and fit its primary initial mass, $M_i$, and ratio of the secondary and primary initial masses, $R_i \leq 1$; a unitary system is expected to have a mass ratio near zero.  Because stellar luminosities sum, and magnitudes are on a log-luminosity scale, the predicted magnitudes for (binary) star $i$ are
\begin{multline}
  \bmu_{i} = 
      -2.5 \log_{10} \Big(10^{-\bG\big(M_i, \boldsymbol{\Theta}, \phi_{Y}\big)/2.5}  \\ + 10^{-\bG\big(M_i R_i, \boldsymbol{\Theta}, \phi_{Y}\big)/2.5}\Big).
      \label{eq:GModel}
\end{multline}
Owing to the nature of the stellar evolution models tabulated in $\bG$, $\bmu_{i}$ is a complex non-linear function of the underlying parameters.

We account for field star contamination by introducing indicator variables $\bZ = (Z_1, \ldots, Z_N)$, where $Z_i = 1$ if star $i$ is a cluster star and $Z_i=0$ if star $i$ is a field star, following the example of \citet{vanDyk_2009}.  These variables allow us to specify a different statistical model for the photometric magnitudes of cluster stars versus those of field stars.  We model the observed photometric magnitudes of cluster stars as $n$-dimensional multivariate Gaussian distributions, such that
\begin{align}
&P(\bX_i | \bSigma_i, M_i,R_i, \bTheta, \phi_{Y}, Z_i=1) =  \nonumber \\
&{1\over \sqrt{(2\pi)^{n}|\bSigma_i|}} \exp\bigg(-{1\over2}\Big(\bX_i - \bmu_i\Big)^\top \bSigma_i^{-1}\Big(\bX_i - \bmu_i\Big)   \bigg). &&
\label{eq:cluster_lik}
\end{align}
While this model appears simple at first glance, it is actually quite complex due to the dependence of $\bmu_i$ on the stellar evolution parameters, and the complex interdependencies therein. That $\bG$ cannot be expressed in closed form yields challenges for inference and computation.

Following \citet{vanDyk_2009}, we specify a simple model for field stars that does not depend on any of the parameters of interest.  Each field star may have its own values for $\thetaage$, $\thetaFeH$ , $\thetamod$, $\thetaabs$, and $\phi_{Y}$, and we cannot fit these parameters.  We therefore simply assume that each field star magnitude is uniformly distributed over the range of the data, such that
\begin{equation*}
P(\bX_i | Z_i=0) = c \quad \hbox{ if } \text{min}_j  \leq x_{ij} \leq \text{max}_j, j=1,\ldots, n,
\end{equation*}
and zero elsewhere, where $(\min_j, \max_j)$ is the range of magnitude values for filter $j$, and $c=\left[\prod_{j=1}^n (\max_j -\min_j)\right]^{-1}$.  We could instead incorporate a more complex and realistic model for field stars; properties of field stars for specific Galactic fields exist and may assist in tuning the model \citep[e.g.,][]{Robin_2012}. However, our work to date has not necessitated the additional effort because the simple model adequately identifies field stars; \changes{This is illustrated using a simulation study in Section~\ref{sec:SimFieldStar}}.


A preliminary likelihood function for a simple stellar population can now be written, 
\begin{eqnarray}
&&L_{\rm p}(\bM,\bR, \bZ, \bTheta, \phi_{Y} | \bX, \bSigma)\nonumber \\
&& \; \; = \prod_{i=1}^{N} 
\Bigg[ 
Z_i \times {1\over \sqrt{(2\pi)^{n}|\bSigma_i|}}
 \exp\bigg(-{1\over2}\Big(\bX_i - \bmu_i\Big)^\top  \nonumber \\ 
&& \qquad \times \bSigma_i^{-1}\Big(\bX_i - \bmu_i\Big)   \bigg) + (1- Z_i) \times P(\bX_i | Z_i=0) \Bigg] \nonumber 
\end{eqnarray}
\begin{eqnarray}
\label{eq:like}
&& \; \; =  \prod_{i=1}^{N} 
\bigg[ 
Z_{i} \times P(\bX_i | \bSigma_i, M_i,R_i, \bTheta, \phi_{Y}, Z_i=1) \nonumber \\
&& \qquad  \qquad  \qquad + (1- Z_i) \times P(\bX_i | Z_i=0)
\bigg],  \\ \nonumber
\end{eqnarray}
where $\bM = (M_1, \ldots, M_N)$, $\bR =(R_1,\ldots, R_N)$, $\bX =(\bX_1,\ldots, \bX_N)$, and $\bSigma = (\bSigma_1,\ldots, \bSigma_N)$.  The sum in (\ref{eq:like}) represents the fact that the sample of stars is a mixture of two subgroups: cluster stars and field stars; such distributions are known is {\it finite mixture distributions} in the statistics literature. \changes{Interested readers are referred to \citet{BayesPhys} for a review of the application of mixture models in astronomy.}

\changes{Rather than embedding $\bG$ into a statistical likelihood function as we do in (\ref{eq:like}), the computer model can be accounted for using a computational approach known as {\it Approximate Bayesian Computation} (ABC). ABC is typically used in situations where the likelihood function is either unavailable or computationally expensive to evaluate, but forward simulation of synthetic data under the statistical model is relatively fast \citep[e.g.,][]{cosmoABC}. While synthetic data can be easily generated under the model in (\ref{eq:like}), constructing a distance measure for comparing observed and synthetic data that accounts for the known Gaussian measurement errors, binary star systems, and field star contamination would be a challenge.}

\subsection{The Likelihood Function for a Two-Population Globular Cluster}
\label{subsec:lik2}

We now extend $P(\bX_i | \bSigma_i, M_i,R_i, \bTheta, \phi_{Y}, Z_i=1)$ in (\ref{eq:cluster_lik}) and (\ref{eq:like}) to account for the fact that the sample of cluster stars is itself a mixture of two subgroups that are the two stellar populations.  This results in a model with three subgroups: field stars and two cluster populations.  The likelihood function for a two-population cluster is then  
\begin{eqnarray}
&&L(\bM,\bR, \bTheta, \bPhi, \bZ | \bX, \bSigma) \nonumber \\
&& \; = \prod_{i=1}^{N} \bigg[ Z_{i} \times \sum_{k=1}^{2}\phi_{pk}P(\bX_i | \bSigma_i, M_i,R_i, \bTheta, \phi_{Yk}, Z_i=1) \nonumber \\
&& \qquad \qquad \qquad + (1- Z_i) \times P(\bX_i | Z_i=0) \bigg].   \label{eq:multipop_like}  \\ \nonumber
\end{eqnarray}
Evaluating (\ref{eq:multipop_like}) involves computing the expected photometry for each star as if it were a member of each population, i.e.,
\begin{eqnarray}
& \bmu_{ik} = -2.5 \log_{10} \Big(10^{-\bG\big(M_i, \boldsymbol{\Theta}, \phi_{Yk}\big)\big/2.5} \nonumber \\
& + 10^{-\bG\big(M_i R_i, \boldsymbol{\Theta}, \phi_{Yk}\big)\big/2.5}\Big), 
\end{eqnarray}
for $i=1,\dots, N$, $k=1,2$.  The population proportions in (\ref{eq:multipop_like}) must sum to one: $\phi_{p1}+ \phi_{p2} = 1$.

\subsection{The Prior Distribution}
\label{subsec:priors}

A key advantage to adopting a Bayesian approach is its ability to directly incorporate previous (independent) results through the joint prior distribution, which we specify via a set of independent priors on each parameter.  For example, we construct a prior distribution on initial mass that is derived from the \citet{MillerScalo_1979} initial mass function.  In particular, we specify a Gaussian prior distribution on the $\textrm{log}_{10}$ of primary initial masses:
\begin{equation}
P\left({\rm log}_{10}(M_i) \right) \propto \text{exp}\left(-\frac{1}{2}\left(\frac{{\rm log}_{10}(M_i) + 1.02}{0.677}\right)^{2}\right),
\end{equation}
truncated to 0.1 $M_{\odot}$ to 8 $M_{\odot}$, where the numerical constants are taken from \citet{MillerScalo_1979}.  For the ratio of the secondary and primary masses we use a uniform prior distribution on $[0,1]$.  We need not truncate the lower end of the secondary mass because low secondary masses indicate that the star is a unitary system \citep{vanDyk_2009}.  

For the cluster parameters, $\bTheta$, we incorporate ancillary information to specify informative (i.e. narrow) prior distributions when available, and use relatively diffuse prior distributions when such information is lacking.  In particular, for $\thetaFeH$, $\thetamod$, and $\thetaabs$, we use Gaussian prior distributions (truncated to be positive in the case of $\thetaabs$), with means set according to previously published values from independent datasets and standard deviations chosen to be reasonably large.  For age, $\thetaage$, we use a uniform prior distribution truncated to the reasonable range of 1 Gyr to 15 Gyr, which includes all Galactic globular clusters.  

Because the population parameters, $\bPhi = (\phi_{Y1},\phi_{Y2},$ $\phi_{p1}, \phi_{p2})$, are the primary parameters of scientific interest, we use uniform prior distributions subject to physical constraints on their ranges.  A uniform prior distribution on the interval $[0.15, 0.3]$ is used for $\phi_{Y1}$; this bounds the helium fraction between 15\% and 30\%.  Similarly, a uniform prior distribution on the interval $[0.15, 0.4]$ is used for $\phi_{Y2}$, and we impose the constraint $\phi_{Y2} > \phi_{Y1}$.  Because we do not typically have prior knowledge for the proportion of stars in each population, $\phi_{p1}$ is given a uniform prior distribution on the interval $[0, 1]$. \changes{When such prior knowledge is available we advocate using a more general beta prior distribution\footnote{A beta$(\alpha,\beta)$ distribution for generic $0 \leq \psi \leq 1$ has the density
\begin{equation}
\frac{\Gamma(\alpha + \beta)}{\Gamma(\alpha)\Gamma(\beta)} \psi^{\alpha - 1} (1 - \psi)^{\beta - 1}, \nonumber
\end{equation}
where $\alpha, \beta > 0$ are shape parameters and $\Gamma(\cdot)$ is the gamma function.
}; a uniform distribution on the interval $[0,1]$ is equivalent to a beta(1,1) distribution}. We do not need to specify a prior distribution for $\phi_{p2}$ because $\phi_{p2} = 1 - \phi_{p1}$.

Ancillary measurements (e.g., proper motions) can be used to probabilistically separate field stars from cluster stars.  When such ancillary measurements are unavailable, we use $P(Z_i=1)=\alpha$ for $i=1,\dots,N$, where $\alpha$ is based on the expected fraction of cluster stars in the dataset.  As we show in Section~\ref{sec:multipop_data_analysis}, our results are not sensitive to reasonable choices of $\alpha$.


\section{Statistical Computation}
\label{sec:MultipopStatComp}

The likelihood function given in (\ref{eq:multipop_like}) and the prior distributions specified in Section~\ref{subsec:priors} complete the model formulation for a two-population globular cluster.  Because our two-population model contains $4$ cluster parameters, $3$ population parameters, and $3 \times N$ stellar parameters, a ``small'' data set containing only $3000$ stars has a parameter space with $9007$ dimensions.  There are also (possibly non-linear) correlations amongst the parameters, see \citet{OMalley_2013}.  The resulting posterior distribution is thus complex and high-dimensional, requiring MCMC techniques for model fitting (see ~\citet{mcmc_handbook} for an overview of MCMC).  MCMC algorithms use an iterative approach to explore the posterior distribution.  In standard MCMC algorithms, again letting $\psi$ represent generic parameters, at iteration $l+1$ new parameter values $\psi^{(l+1)}$ are generated from a distribution $\Gamma$ that depends only on the data and the current parameter values $\psi^{(l)}$.  After $L$ iterations, MCMC produces a correlated sample of parameter values, $\{\psi^{(1)}, \dots, \psi^{(L)} \}$, known as an {\it MCMC chain}.  

With an appropriate choice of $\Gamma$ and after a sufficient number of iterations, known as {\it burn-in}, the chain converges to a {\it stationary distribution} and the MCMC sample can be regarded as a (correlated) sample from $P(\psi | Y)$. A popular method of obtaining $\Gamma$ is the {\it Metropolis algorithm} \citep{Metropolis_1953}.  After drawing $\psi^{(1)}$ from some starting distribution, the Metropolis algorithm consists of two-steps.  For iterations $l=2,\dots, L$:

\begin{enumerate}

\item Draw a ``proposed state'' $\psi^{(*)}$ from a proposal distribution that is symmetric about $\psi^{(l-1)}$ (e.g., a Gaussian distribution centered at $\psi^{(l-1)}$).

\item With probability $\text{\rm min} \Big( 1,  \frac{P(\psi^{(*)} | Y)}{P(\psi^{(l-1)} | Y)} \Big)$, set $\psi^{(l)} =  \psi^{(*)}$.  Otherwise, set $\psi^{(l)} =  \psi^{(l-1)}$.

\end{enumerate}

The efficiency of the Metropolis algorithm depends heavily on the choice of proposal distribution in the first step.  If the distribution is too narrow, many proposed $\psi^{(*)}$ are accepted (i.e.,  $\psi^{(l)}$ is set to $\psi^{(*)}$ in the second step) but MCMC takes small steps.  Consequently, the chain may take a long time to converge to the posterior distribution and $\{\xi^{(1)}, \dots, \xi^{(L)} \}$ will have high autocorrelation.  Conversely, if the proposal distribution is too wide, there will be a few big steps, but many rejected $\psi^{(*)}$.  When this happens, the chain can become stuck at a particular parameter value for many iterations and not fully explore the posterior distribution.  A good choice of proposal distribution is generally non-obvious and requires either fine-tuning or more sophisticated approaches.

Our MCMC strategy for fitting the two-population model relies on two key techniques: marginalization and adaptation.  Complex posterior correlations and multiple modes frustrate convergence of MCMC.  By marginalizing over (i.e., integrating out) the stellar parameters, an approach initially devised by \citet{Stein_2013}, we lessen multi-modality and dramatically reduce the dimension of the posterior distribution from $9007$ to $7$ for a data set with $3000$ stars.  Adapting the proposal distribution to the resulting (marginal) posterior distribution further improves efficiency (compared to a standard Metropolis algorithm).  We discuss marginalization and adaptation in Sections~\ref{sec:Marg} and \ref{sec:adapMCMC}, respectively.

\subsection{Marginalization via Numerical Integration}
\label{sec:Marg}

With the full joint posterior distribution denoted by $P(\bTheta, \bPhi, \bM, \bR, \bZ \mid \bX)$, the marginal posterior distribution of $(\bTheta, \bPhi)$ is given by
\begin{align}
	&P(\bTheta, \bPhi \mid \bX) \nonumber \\
	& = \int \cdots \int \Big(\sum_{Z_1} \cdots \sum_{Z_N} 
	P(\bTheta,  \bPhi, \bM, \bR, \bZ \mid \bX)\Big) d\bM d\bR \label{multi:margpost1} \\
	& \propto P(\bTheta, \bPhi) \prod_{i=1}^N
	\bigg[ c(\bTheta, \bPhi)P(Z_i=1) \nonumber \\
	& \qquad\qquad\qquad\qquad\qquad + P(\bX_i | Z_i=0)P(Z_i=0) \bigg], \label{multi:margpost}
\end{align}
where $c(\bTheta, \bPhi)$
\begin{align}
	& = \int \int \sum_{k=1}^{2}\phi_{pk}P(\bX_i | \bSigma_i, M_i,R_i, \bTheta, \phi_{Yk}, Z_i=1) \nonumber \\
	& \qquad\qquad\qquad\qquad \times P(M_i, R_i \mid Z_i=1) dM_i dR_i.
\end{align}
When $P(Z_i=1) = \alpha$ for $i=1,\dots,N$ (i.e., when all stars have the same probability of being cluster stars), (\ref{multi:margpost1}) and (\ref{multi:margpost}) reduce to $P(\bTheta, \bPhi \mid \bX)\propto$
\begin{align}
& P(\bTheta,\bPhi) \prod_{i=1}^{N} \bigg[ (1- \alpha)  P(\bX_i | Z_i=0) \nonumber \\
& \qquad + \int \int  \alpha \sum_{k=1}^{2}\phi_{pk}P(\bX_i | \bSigma_i, M_i,R_i, \bTheta, \phi_{Yk}, Z_i=1) \nonumber \\
& \qquad \qquad \times P(M_i, R_i \mid Z_i=1) dM_i dR_i \bigg]. 
\label{eq:multipop_post}
\end{align}
This integral cannot be evaluated analytically because $P(\bX_i \mid M_i, R_i, \bTheta, \phi_{Yk}, Z_i=1)$ depends on $M_i$ and $R_i$ through $\bG$ \citep{Stein_2013}.  Instead, we employ brute-force numerical integration via Riemann sums.  By marginalizing out the $3N$ stellar parameters we reduce the dimension of the posterior distribution from typically thousands to just seven.    

\subsection{Adaptive MCMC}
\label{sec:adapMCMC}

Because the remaining parameter vector $(\bTheta, \bPhi)$ after marginalizing out $\bM$, $\bR$, and $\bZ$ is just seven-dimensional, we initially implemented a standard Metropolis algorithm to sample from $P(\bTheta, \bPhi | \bX)$.  However, we found the trial-and-error approach to tuning the (seven-dimensional) proposal distribution to be difficult; this is not surprising given the correlations among the components of $(\bTheta, \bPhi)$ in $\bG$.  To avoid arduous fine-tuning and make BASE-9 more accessible to users less familiar with MCMC, we implement an {\it Adaptive Metropolis} (AM) algorithm \citep[e.g.,][]{Haario_2001, Roberts_2009, Rosenthal_2011}.  Whereas an iteration of a standard Metropolis algorithm only depends on the most recent value in the MCMC chain, an AM algorithm uses the entire history of the chain to adapt the proposal distribution at each iteration.  This, however, violates a defining property of a Markov chain: the distribution of a value in the chain can only depend on the history of the chain through its most recent value.  Thus, care must be taken to guarantee an AM algorithm converges properly.  As recounted in \citet{Rosenthal_2011}, AM algorithms must satisfy the {\it Diminishing Adaptation Condition}: the amount of adaptation at iteration $l$ must go to $0$ as $l \rightarrow \infty$.  The Diminishing Adaptation Condition is key; other technical conditions are almost always satisfied except in specially constructed examples \citep{Rosenthal_2011}.  Readers interested in additional mathematical details are encouraged to consult \citet{Rosenthal_2011} and references therein.  

\begin{figure*}[ht!]
\begin{center}
\includegraphics[width=\textwidth]{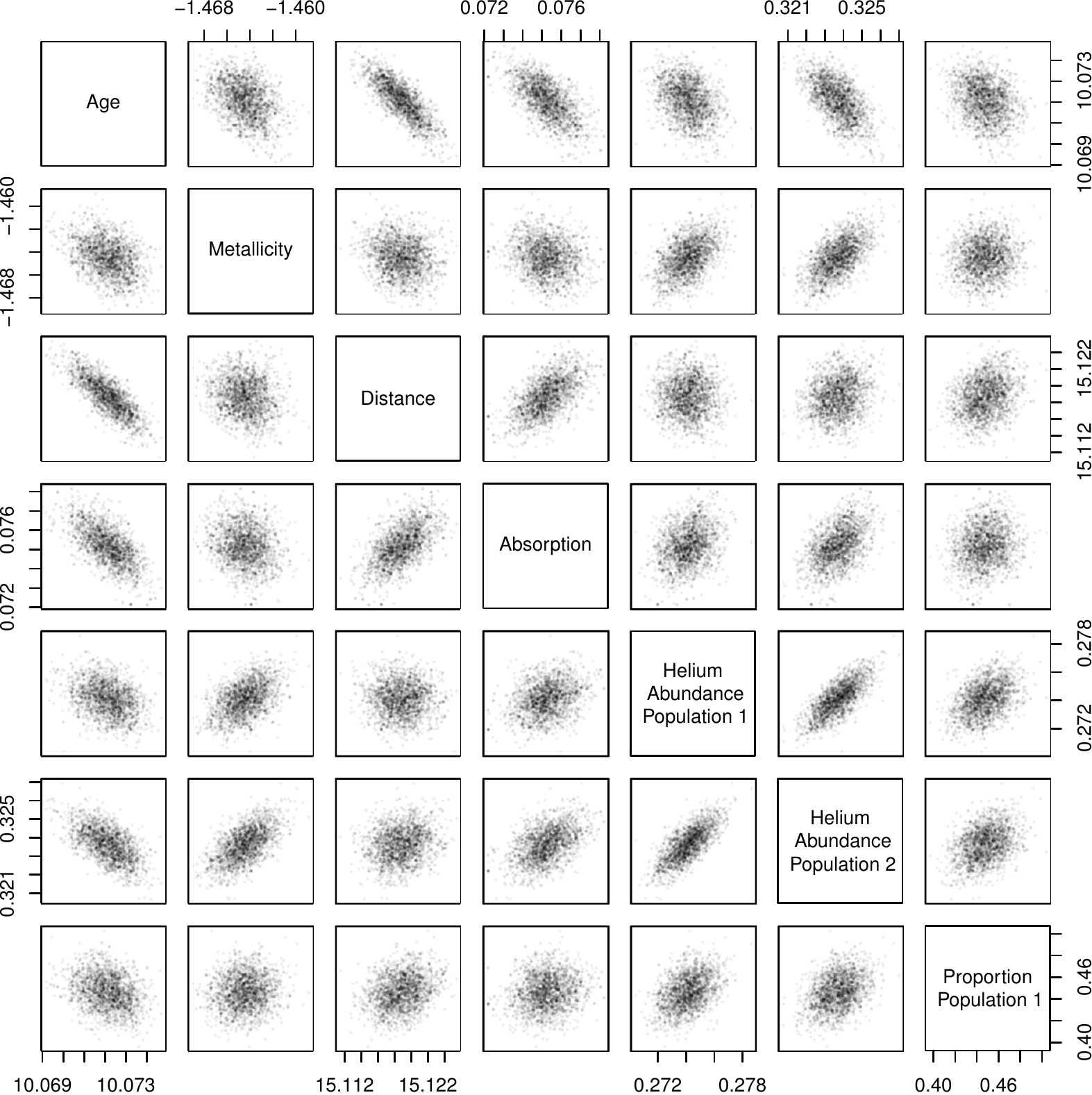}
\end{center}
\caption[Posterior draws from the marginal posterior distribution $P(\bTheta, \bPhi | \bX)$] {Posterior draws from the marginal posterior distribution $P(\bTheta, \bPhi | \bX)$.  The model was fit using photometric magnitudes from NGC 5272.  From these draws, $P(\bTheta, \bPhi | \bX)$ appears roughly Gaussian.}
\label{fig:ScatterPlot}
\end{figure*}

After marginalizing over $\bM$, $\bR$, and $\bZ$, the resulting marginal posterior distribution $P(\bTheta, \bPhi | \bX)$ given in (\ref{eq:multipop_post}) appears roughly Gaussian.  This is illustrated in Figure~\ref{fig:ScatterPlot}, which displays the matrix of two-dimensional scatterplots of 25,000 posterior draws from $P(\bTheta, \bPhi | \bX)$.  (The data we used to construct these plots are photometric magnitudes from NGC 5272.  Details are provided in Section~\ref{sec:multipop_data_analysis}.)  Based on results in \citet{Gelman_1996}, the optimal proposal distribution for a Gaussian posterior distribution with a $d$-dimensional covariance matrix $\Upsilon$ is itself a Gaussian distribution with covariance matrix $\left[ \left(2.38\right)^2/d \right]\Upsilon$.  Because the actual form of the posterior distribution is unknown, for iteration $l+1$ we use a multivariate $t$ proposal distribution with 6 degrees of freedom\footnote{For generic parameters $\Psi$ with $p$-dimensional scale matrix $\Omega$, at iteration $l+1$ the multivariate $t$ proposal distribution with $\nu$ degrees of freedom has the density
\tiny
\begin{equation}
\frac{\Gamma[(\nu + p)/2]}{\Gamma(\nu/2)\nu^{p/2}\pi^{p/2} |\Omega|^{1/2}[1+ \frac{1}{\nu}(\Psi^{(l+1)} - \Psi^{(l)})^\top \Omega^{-1}(\Psi^{(l+1)}-\Psi^{(l)})]^{(\nu+p)/2} \nonumber}.
\end{equation}
\footnotesize
The multivariate $t$ distribution has a similar ``bell shape'' to the multivariate Gaussian distribution, but with fatter tails.
}
, centered at the current value of $\left(\bTheta,\bPhi\right)$ and with scale equal to $\left[ \left(2.38\right)^2/7 \right]\bxi^{(l)}$.  Here, $\bxi^{(l)}$ is the empirical variance-covariance matrix of $\{ (\bTheta^{(1)}, \bPhi^{(1)}), \dots, (\bTheta^{(l)}, \bPhi^{(l)})\}$.  Because we recalculate $\bxi^{(l)}$ at every iteration, the proposal distribution adapts at every iteration based on the past history of the chain.  As $l \rightarrow \infty$, the empirical distribution of $\{ (\bTheta^{(1)}, \bPhi^{(1)}), \dots, (\bTheta^{(l)}, \bPhi^{(l)})\}$ approaches the marginal posterior distribution $P(\bTheta, \bPhi | \bX)$, improving efficiency.  Furthermore, $\bxi^{(l)}$ stabilizes and thus the adaptation diminishes \changes{as required}.

\changes{Alternative modern MCMC approaches include Hamiltonian Monte Carlo (see, e.g., \url{http://mc-stan.org}) and Riemann manifold Monte Carlo methods \citep{Girolami2011}. Such methods are particularly useful when the posterior distribution exhibits strong correlations and curving degeneracies. Hamiltonian Monte Carlo (HMC), for example, borrows ideas from Hamiltonian dynamics to make sampling more efficient, but typically requires the likelihood to be available analytically and that its derivatives with respect to the model parameters be available. Similar caveats apply to Riemann manifold Monte Carlo (RMMC). While we could develop an analytical emulator of the function $\bG$ to deploy HMC or RMMC, the additional effort is unnecessary due to the roughly Gaussian shape of $P(\bTheta, \bPhi | \bX)$; the AM algorithm automatically improves sampling efficiency by adapting the proposal distribution.}

When implementing the AM algorithm, we first run the sampler in ``tuning'' mode.  The goal of this tuning period is not to obtain an optimal proposal distribution, but rather to sufficiently explore the posterior distribution and generate a reasonable $\bxi^{(1)}$ for the AM algorithm (see the Appendix for details).    Once we have calculated $\bxi^{(1)}$ from the tuning period, the first 1000 iterations of the AM algorithm use the multivariate $t$ proposal distribution described above, with $\bxi^{(l)}=\bxi^{(1)}$ for $l=1,\dots,1000$.  This non-adaptive period is necessary to generate a sufficiently large sample to estimate posterior covariances before adapting the proposal distribution. At iteration 1001 and at every subsequent iteration, $\bxi^{(l)}$ is the empirical covariance matrix of the previous $l$ iterations.

\begin{figure*}[ht!]
\begin{center}
\includegraphics[width=0.925\textwidth]{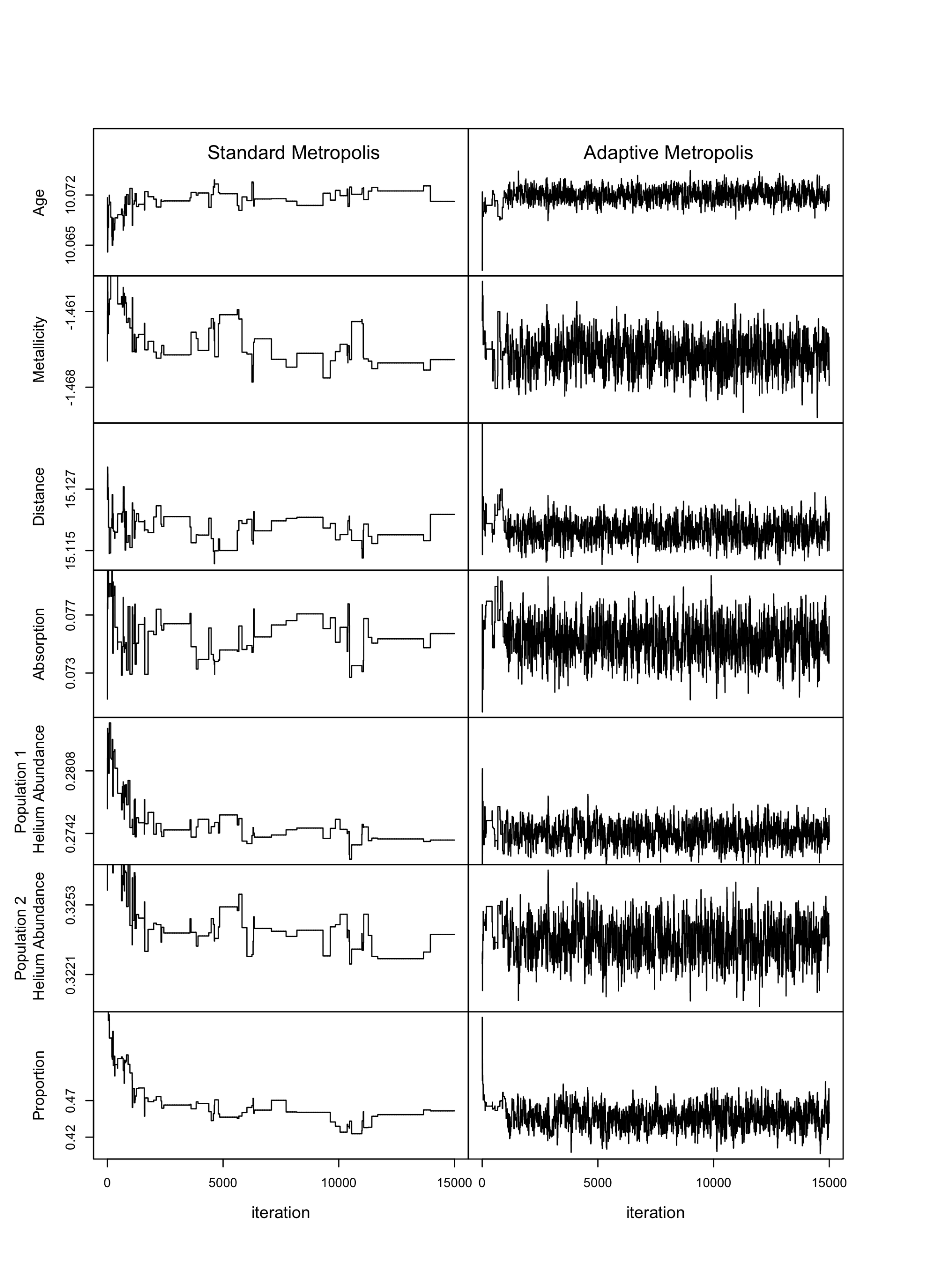}
\end{center}
\vspace{-0.5cm}
\caption[Improving convergence with an Adaptive Metropolis algorithm] {Improving convergence with an Adaptive Metropolis algorithm.  The left column presents trace plots for a non-adaptive Metropolis algorithm, and the right column presents the trace plots for the Adaptive Metropolis algorithm we devised.  Both algorithms used the same data and are implemented with the same starting values and tuning procedure.}
\label{fig:AdapMetrop}
\end{figure*}

The efficiency of our AM algorithm is demonstrated in Figure~\ref{fig:AdapMetrop}.  There, we compare the performance of our AM algorithm to that of a standard Metropolis algorithm for sampling the same posterior distribution.  The standard Metropolis sampler is identical to the AM sampler except $\bxi^{(l)}$ is fixed at $\bxi^{(1)}$ throughout.  Both algorithms are implemented with the same starting values, the same tuning period, and use the same data as in Figure~\ref{fig:ScatterPlot}.  The proposal distribution in the AM algorithm begins adapting at iteration 1001, after which there is an obvious difference in performance between AM and standard Metropolis.  Standard Metropolis struggles as the MCMC chain repeatedly becomes stuck.  While AM also sticks initially, adapting the proposal distribution quickly frees the chain and leads to increased efficiency.  As expected, the AM algorithm becomes increasingly efficient as the number of iterations increases.

\section{Simulation Studies}
\label{sec:MultipopNumRes}

\subsection{Recovering Two Population Clusters}
\label{sec:SimTwoPop}

\begin{figure}[t!]
\begin{center}
\includegraphics[width=0.475\textwidth]{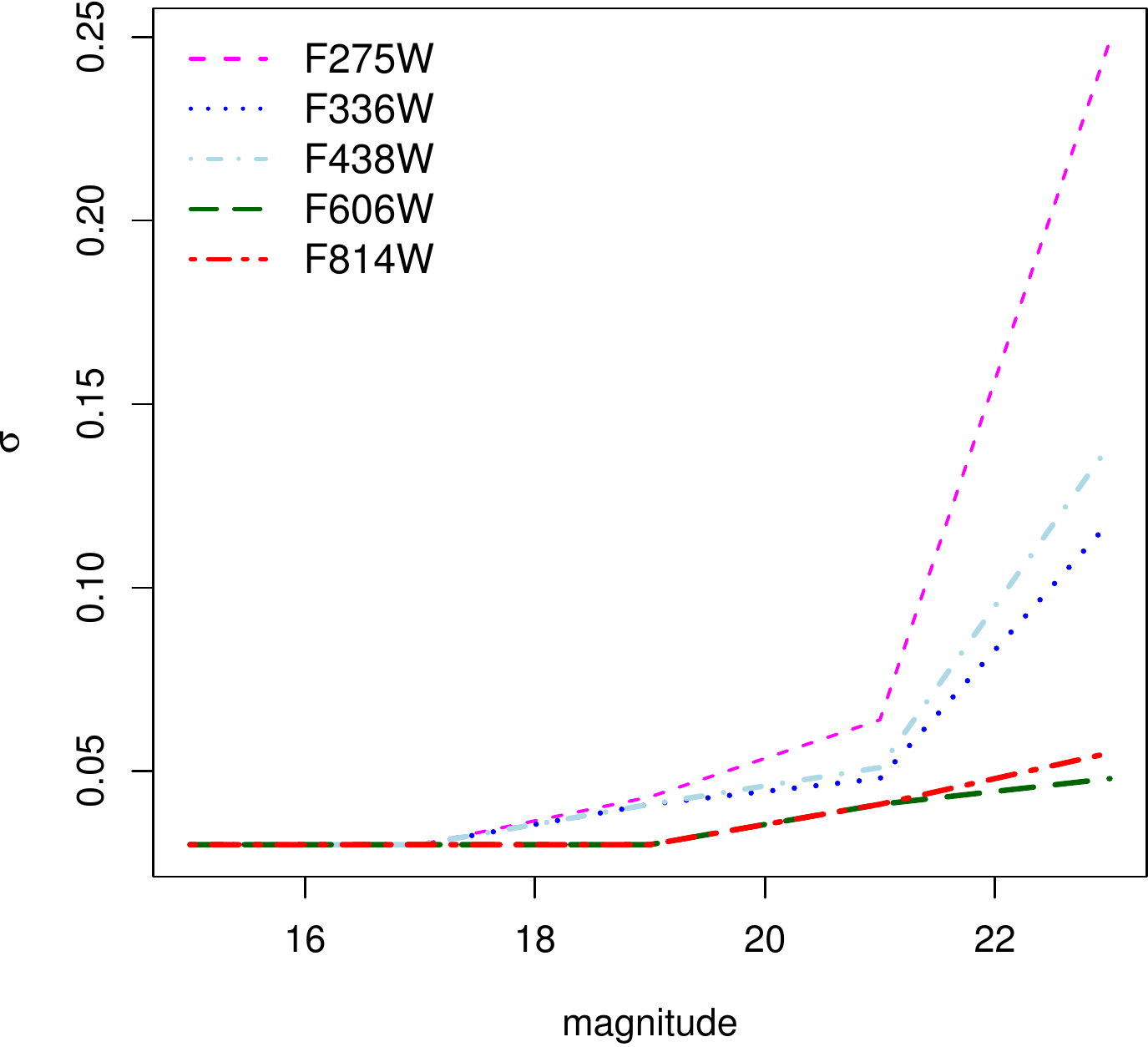}
\end{center}
\caption[Gaussian measurement error for simulated two-population clusters] {Gaussian measurement error for simulated two-population clusters.  In our numerical studies, the photometric magnitudes for each star are simulated with Gaussian measurement error that is representative of the measurement error we expect for observed data, which is depicted above.  Here, $\sigma$ is the standard deviation of the Gaussian measurement error.}
\label{fig:meas_error}
\end{figure}

As an initial test of our method, we simulate two-population globular clusters under three scenarios, with ten replicate clusters per scenario.  The three scenarios differ in the percentage of stars belonging to Population 1: 50\%, 80\%, and 100\% for scenario 1, 2, and 3, respectively.  Scenario 3 therefore contains ten replicates of a single-population cluster, which we intentionally fit with our (incorrect) two-population model to demonstrate how model misspecification can \changes{potentially be identified}.  Each cluster is simulated with $\thetaage = 10.08$, $\thetamod = 15.375$, $\thetaabs = 0.372$, and $\thetaFeH = -1.5$, which are ``average'' published values across the clusters compiled in \citet[][and updated in 2010 at \href{http://physwww.mcmaster.ca/~harris/mwgc.ref}{http://physwww.mcmaster.ca/$\sim$harris/mwgc.ref}]{Harris_1996}.  In the simulations, we set $\phiY_1 = 0.24$ and $\phiY_2 = 0.29$, so that the true difference in helium abundance is 0.05.  We simulate 30,000 cluster stars and 1000 field stars per cluster, and every star is generated as a single-star system.  (Future work will include binaries.)  For each cluster we generate photometric magnitudes in five filters, corresponding to the filters in HST UVIS photometry \citep{Piotto_2015}: $F275W$, $F336W$, $F438W$, $F606W$, and $F814W$.  Details about these filters are provided in Section~\ref{sec:multipop_data_analysis}.  The photometric magnitudes for each star are simulated with uncorrelated Gaussian measurement error that is a function of both the wavelength band and the magnitude, as depicted in Figure~\ref{fig:meas_error}.   

Of the 31,000 stars generated per cluster, about 90\% are dropped from the simulated cluster for one of two reasons.  First, there is a threshold signal-to-noise ratio that eliminates stars too dim to be observed under realistic conditions.  Second, we believe the stellar evolution models are inaccurate for fainter main sequence stars and we therefore impose a magnitude cutoff on real photometry \citep{vanDyk_2009, DeGennaro_2009}.  We impose the same cutoff on simulated photometry so that our simulation results are as informative as possible.  The exact cutoff we use depends on the assumed distance to the cluster (see Section~\ref{sec:multipop_data_analysis}).  In the simulations, we discard stars with a photometric magnitude in the $F275W$ filter greater than 23.  After losing stars due to low signal-to-noise and the $F275W$ magnitude cutoff, about 3000 simulated stars remain per cluster.

We use prior distributions for $\thetamod$, $\thetaabs$, and $\thetaFeH$ with means equal to their true values under the simulation; the prior standard deviations were set to 0.05, 0.124, and 0.05, respectively.  The prior distributions on the population parameters are as described in Section~\ref{subsec:priors}.  We assign $P(Z_i=1)=\alpha=0.95$ for $i=1,\dots,N$.  This is the value for $\alpha$ we use when analyzing NGC 5272 in Section~\ref{sec:multipop_data_analysis} after testing the sensitivity to the choice of $\alpha$.  

\begin{figure*}[p!]
\begin{center}
{\normalsize (a) Scenario 1: 50\% of stars in Population 1} \\
\vspace{0.025cm}
\includegraphics[width=\textwidth]{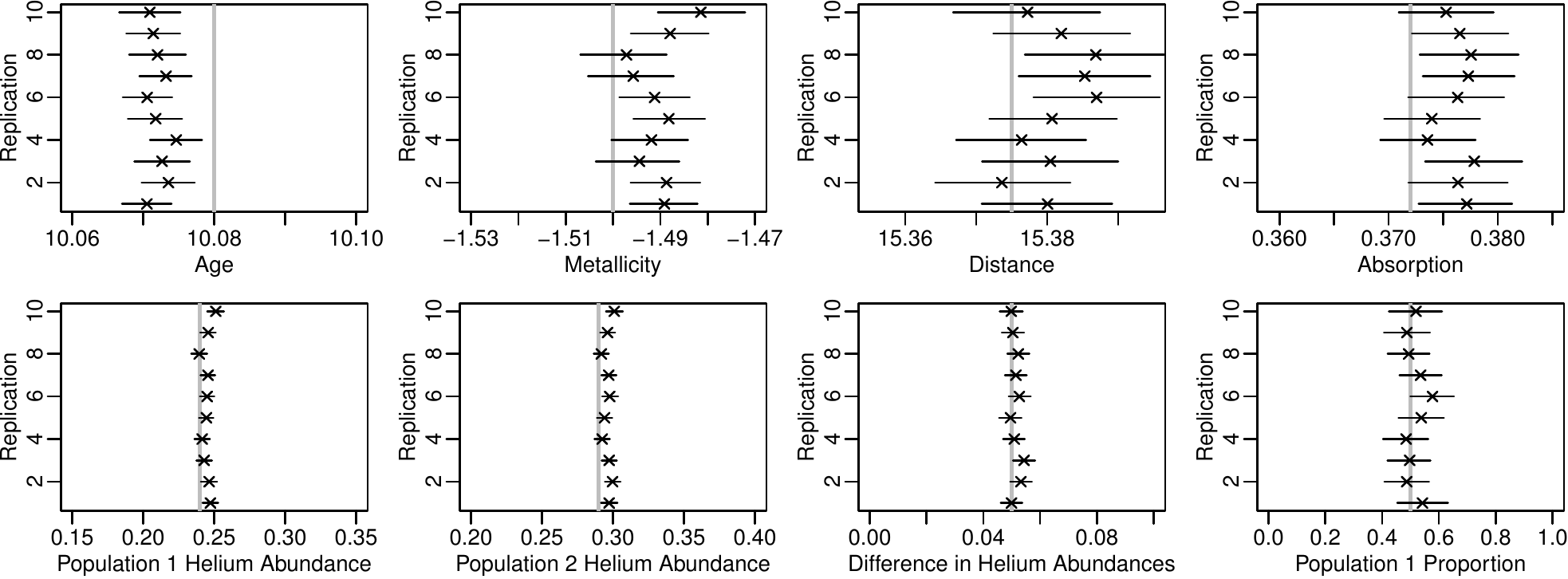} \\
\vspace{0.5cm}
{\normalsize (b) Scenario 2: 80\% of stars in Population 1} \\
\vspace{0.025cm}
\includegraphics[width=\textwidth]{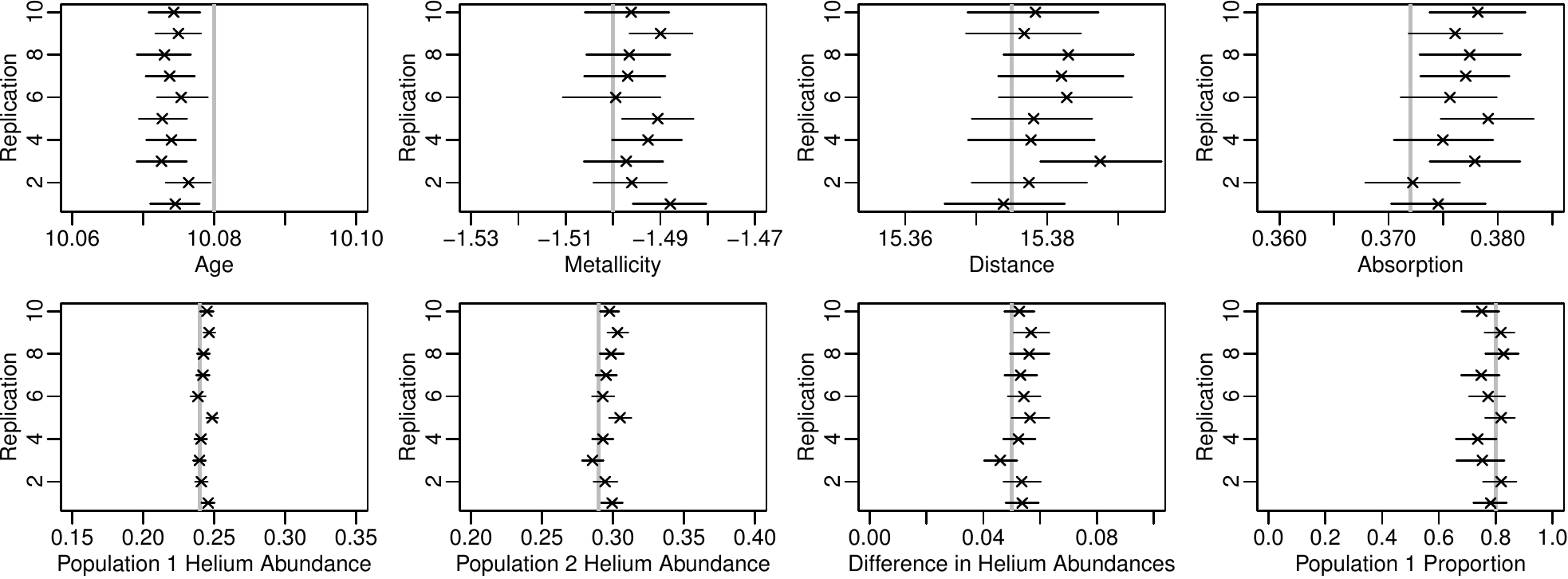} \\ 
\vspace{0.5cm}
{\normalsize (c) Scenario 3: 100\% of stars in ``Population 1''} \\
\vspace{0.025cm}
\includegraphics[width=\textwidth]{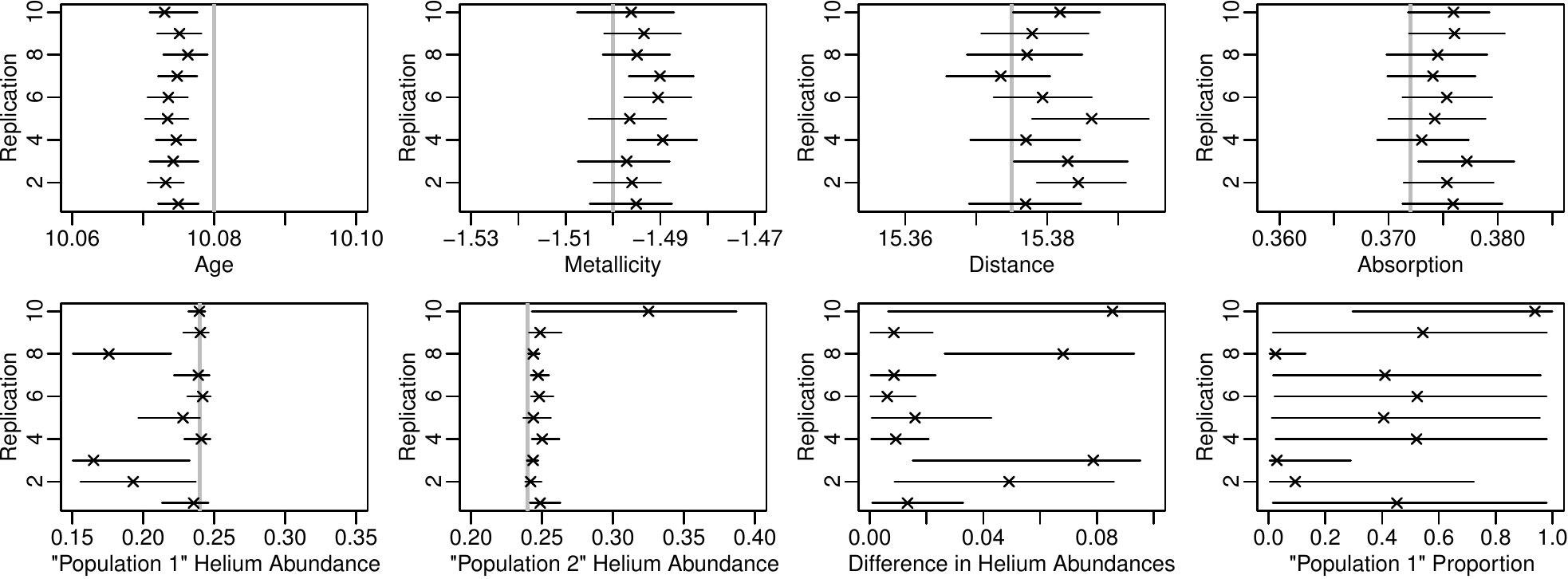} \\ 
\end{center}
\caption{Results of the simulation study.  The horizontal bars are 95\% posterior intervals, with posterior means marked by an `x'.  The true parameter values under the simulation are indicated by the grey vertical lines.}
\label{fig:numerical_study_res}
\end{figure*}

We use our AM algorithm to explore $P(\bTheta, \bPhi | \bX)$ for each of the thirty simulated clusters.  We run one chain per cluster for 25,000 iterations after the tuning period.  Inspection of the trace plot for each chain shows that all the chains reach their apparent stationary distributions within the first 5000 iterations.  We discard the first 5000 iterations of each chain as burn-in and base inference on the remaining 20,000 iterations.  Results for the three scenarios appear in Figure~\ref{fig:numerical_study_res}.

The results for scenarios 1 and 2 are presented in the top four rows of Figure~\ref{sec:multipop_data_analysis}.  There, we observe that our method is performing reasonably well with respect to recovering the difference in helium abundance and the proportion of stars in each population.  This is encouraging, as our main inferential goal is to recover the difference in helium abundance.  Unfortunately, there is a systematic difference between the fitted parameters and the true values of the parameters under the simulation.  The reasons for this discrepancy are examined and discussed in detail in Section 4.3 of \citet{Stenning_PhD}.  It was discovered that the deviations increase with the size of the measurement errors, suggesting an influence of the prior distribution.  The cause is that as the sample size increases, the influence of the prior distribution on primary initial mass does not diminish because there is only one observation (i.e. one star) per mass parameter.  Future work will focus on fitting the distribution of the masses to hopefully eliminate the deviations.  For now, we simply note that the systematic deviations are small relative to both the systematic errors stemming from the underlying stellar evolution model and to the best available statistical errors on these parameters using other methods.  For example, minimum star-by-star [Fe/H] statistical and systematic errors are approximately 0.015 and 0.03 dex, respectively \citep{Carretta_2009b}.  Typical statistical errors for distance moduli are $\sigma(m-M_V)$ = 0.1 mag and those for absorption are $\sigma$($A_{\rm V}$) = 0.1$A_{\rm V}$, with a lower limit of 0.03 mag \citep[][and as updated at \href{http://physwww.mcmaster.ca/~harris/mwgc.ref}{http://physwww.mcmaster.ca/$\sim$harris/mwgc.ref}]{Harris_1996}.  Furthermore, we can adequately recover the relative difference in helium abundance because the systematic differences are in the same direction and to similar degree for both populations.

To check that the recovered helium abundance difference is due to the presence of two populations and is not an artifact of the method, in scenario 3 we intentionally fit simulated single-population clusters with the (now incorrect) two-population model.  The results for the cluster parameters are similar to those in scenarios 1 and 2; see row 5 of Figure~\ref{sec:multipop_data_analysis}.  This is expected because the cluster parameters are common to both populations.  However, in the last row of Figure~\ref{sec:multipop_data_analysis}, the results for the proportion of stars in Population 1 indicate model misspecification.  Specifically, we observe that the fitted value is close to either zero or one (replicates 2, 3, 8, and 10) and/or the 95\% interval is very wide, spanning most of the range from $[0,1]$ (replicates 1, 2, 4, 5, 6, 7, 9, and 10).  Both of these outcomes suggest that a second population may not be present in the data. 

\changes{We caution that investigating intervals and estimates in this way does not provide a formal diagnostic for model misspecification but results such as those under scenario 3 should be considered as ``smoking-gun'' evidence that the two-population model has been applied to a single-population cluster. For now, we intend our model to be used in cases where there are two prominent populations as viewed in CMDs; e.g., two populations for NGC 5272 can be seen in the rightmost CMD in Figure~\ref{fig:MultiPop_CMDs}. Formal criteria to infer the number of populations in a cluster will be included when our model and algorithms are extended to accommodate three or more populations.}

\changes{
\subsection{Testing the Field Star Model}
\label{sec:SimFieldStar}

To test the adequacy of using the simple uniform model described in Section~\ref{subsec:lik1} for field star magnitudes, we simulate five replicate single-population clusters with parameters equal to those reported for NGC 5272 in the updated \citet{Harris_1996} globular cluster catalogue. Field stars are simulated from the Besan\c{c}on model \citep{FieldStarModel} with Galactic $l, b$ = 42.2170, +78.7069, though we increase the field size $100\times$ (from 0.0013 to 0.130 square degrees) relative to that for NGC 5272 to provide an ample sample in each replication. After removing stars due to low signal-to-noise and an imposed magnitude cutoff at $F275W=22.074$ (see Section~\ref{sec:multipop_data_analysis} for a discussion regarding our choice of cutoff), there are approximately 2100 cluster stars and 110 field stars per replicate data set. Using BASE-9, we are able to infer the posterior probability that a star is a cluster member; see \citet{Stein_2013}. Those stars with greater than 50\% posterior probability are classified as cluster stars. We can evaluate the resulting classification using the confusion matrices in Table~\ref{tbl:fieldstar}. A confusion matrix is a table with columns representing true classifications and rows representing predicted classifications. For example, the confusion matrix for Replication 1 reveals that 111 field stars are correctly identified as such, while 1 field star is misclassified as a cluster star; all 2103 clusters stars in Replication 1 are correctly classified. In this simulation, the simple model for field star magnitudes misidentifies field stars as cluster stars $< 2\%$ of the time and never misidentifies cluster stars as field stars. Based on these results, a more complex model for field stars seems unnecessary.

\begin{table}[t!]
\begin{center}
\caption{Confusion Matrices for Cluster Member vs. Field Star}
\vspace{0.2cm}
\label{tbl:fieldstar}
\begin{tabular}{lrcc}
\hline
\multicolumn{4}{c}{{\bf Replication 1}}  \\ \hline \vspace{0.075cm}
& & \multicolumn{2}{c}{\it Observed} \\
& & Field Star & Cluster Member \\	
\multirow{2}{*}{\it Predicted} & Field Star & 111 & 0 \\
& Cluster Member & 1 & 2103 \vspace{0.15cm} \\
\hline
\multicolumn{4}{c}{{\bf Replication 2}}  \\ \hline \vspace{0.075cm}
& & \multicolumn{2}{c}{\it Observed} \\
& & Field Star & Cluster Member \\	
\multirow{2}{*}{\it Predicted} & Field Star & 110 & 0 \\
& Cluster Member & 0 & 2105 \vspace{0.15cm} \\
\hline
\multicolumn{4}{c}{{\bf Replication 3}}  \\ \hline \vspace{0.075cm}
& & \multicolumn{2}{c}{\it Observed} \\
& & Field Star & Cluster Member \\	
\multirow{2}{*}{\it Predicted} & Field Star & 109 & 0 \\
& Cluster Member & 2 & 2080 \vspace{0.15cm} \\
\hline
\multicolumn{4}{c}{{\bf Replication 4}}  \\ \hline \vspace{0.075cm}
& & \multicolumn{2}{c}{\it Observed} \\
& & Field Star & Cluster Member \\	
\multirow{2}{*}{\it Predicted} & Field Star & 109 & 0 \\
& Cluster Member & 2 & 2130 \vspace{0.15cm} \\
\hline
\multicolumn{4}{c}{{\bf Replication 5}}  \\ \hline \vspace{0.075cm}
& & \multicolumn{2}{c}{\it Observed} \\
& & Field Star & Cluster Member \\	
\multirow{2}{*}{\it Predicted} & Field Star & 111 & 0 \\
& Cluster Member & 1 & 2125 \vspace{0.15cm} \\
\hline

\end{tabular}
\end{center}
\end{table}
}


\section{Analysis of NGC 5272}
\label{sec:multipop_data_analysis}

In this section we apply our method to photometric observations of NGC 5272 in order to provide a proof of concept.  Our main objective is to estimate the difference in helium abundance between the two postulated stellar populations, as well as the proportion of stars in each.  A secondary objective is to evaluate the underlying stellar evolution model by examining how well the fitted models agree with the observed data.  We are of course also interested in estimating the other cluster parameters.  The observed data are HST UVIS photometry \citep{Piotto_2015} in five filters per star: $F275W$, $F336W$, $F438W$, $F606W$, and $F814W$.  A detailed description of the data collection and processing appears in \citet{Piotto_2015}.

\begin{figure*}[ht!]
\includegraphics[width=0.5\textwidth]{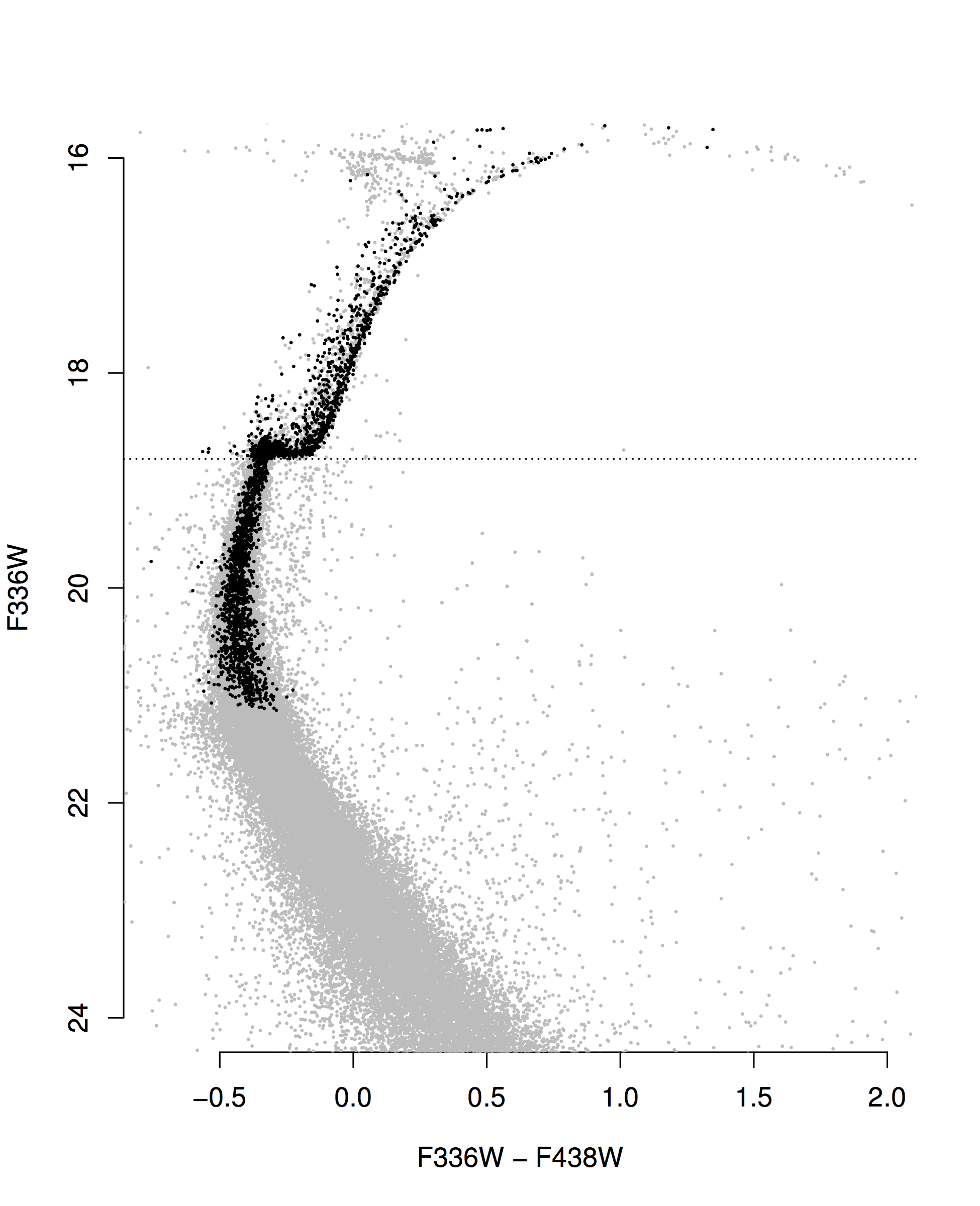}
\includegraphics[width=0.5\textwidth]{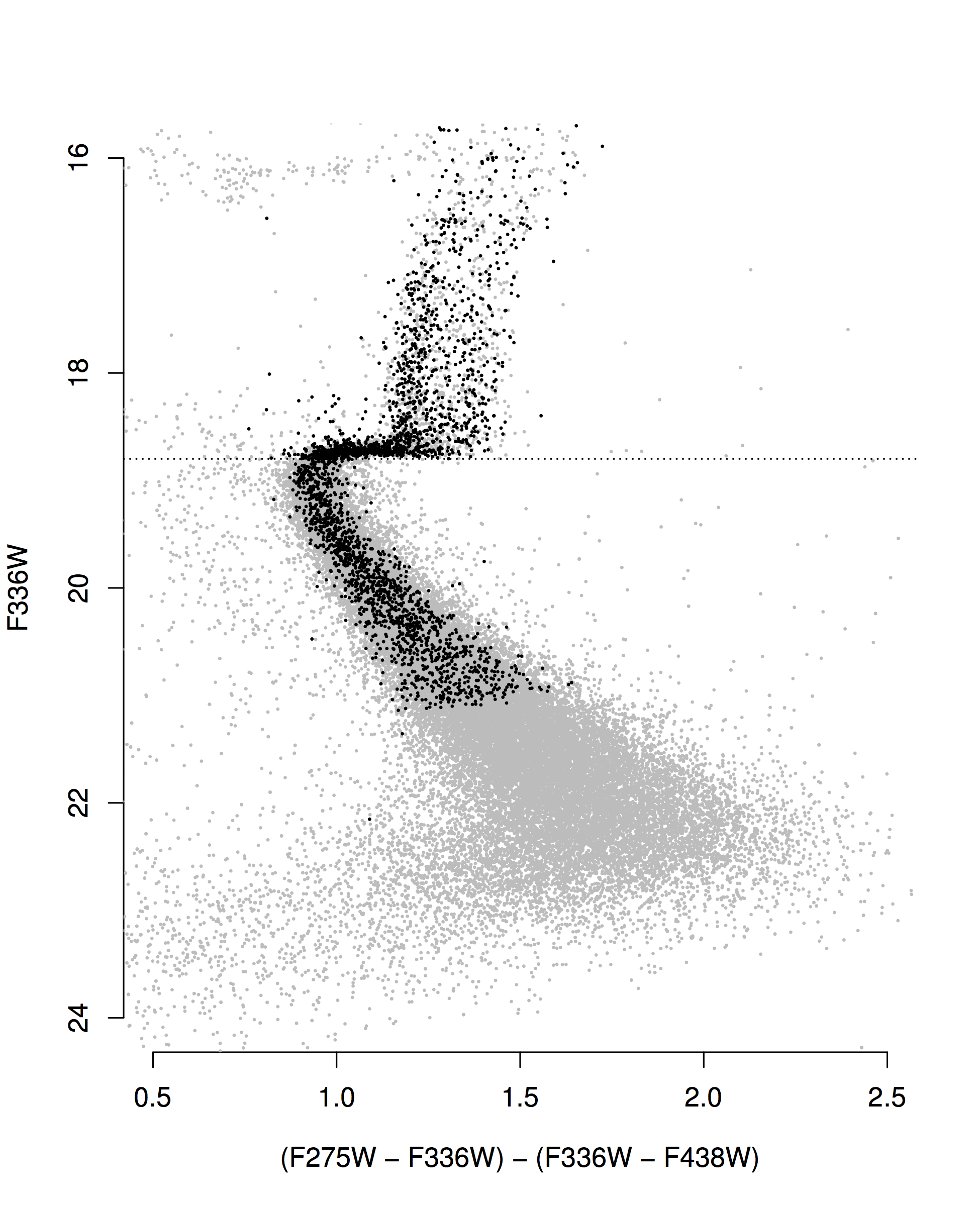}
\caption{\changes{Two CMDs for NGC 5272.  Stars used in our analysis are represented in black; those removed are in grey.  The horizontal dotted lines indicate the cutoff we use to separately sample main sequence and post-main-sequence stars. The CMD on the right uses a combination of three UV filters on the horizontal axis to better display the two populations.}}
\label{fig:MultiPop_CMDs}
\end{figure*}

The data for NGC 5272 consists of 179,330 observed stars; its CMD appears in Figure~\ref{fig:MultiPop_CMDs}.  Because fitting our two-population model with this amount of data is currently computationally impractical, we reduce the observed data.  The stars that remain after reduction are indicated with black dots in the \changes{CMDs} in Figure~\ref{fig:MultiPop_CMDs}; discarded stars are indicated with grey dots.  Our data-reduction routine proceeds as follows:

\begin{enumerate}

\item Pixel location errors are used to remove stars that are likely field stars and quality flags are used to remove stars with poor photometry.

\item By examining the CMD we make general cuts to remove some horizontal branch stars because stellar evolution models for this transitionary phase are not included among our current set of stellar evolution models.  

\item Because we believe that the computer models are particularly inaccurate for the faintest stars, we impose a magnitude cutoff of $F275W=22.074$.  The cutoff is set at $M_V=7$, based on the distance modulus for the cluster reported in the updated \citet{Harris_1996} globular cluster catalogue.  We use an absolute magnitude-based cutoff to enable consistency in future analyses of different clusters.

\item We sample from the remaining stars so that the final photometry set contains 3000 stars.  To do this, we visually identify the main sequence turn off and choose a magnitude cut point to separate main sequence from post-main-sequence stars.  In doing so, we err on the side of including (nearly) all post-main-sequence stars above the cutoff.  For NGC 5272, the cutoff is at $F336W=18.8$, which we indicate with the horizontal dotted line in Figure~\ref{fig:MultiPop_CMDs}.  We sample 1500 stars each from above and below the cutoff, such that our final photometry set contains an equal mix of main sequence and post-main-sequence stars.

\end{enumerate}

Because accurate photometric errors are not yet available for this UV photometric dataset, we construct approximate errors using the HST exposure time calculator and adopt a conservative minimum error of 0.01 magnitude.  As with our simulated clusters in Section~\ref{sec:MultipopNumRes}, the errors are a function of both filter and wavelength.  Additional discussion is provided in \citet{WagnerKaiser_2015}.

\begin{figure*}[ht!]
\includegraphics[width=0.95\textwidth]{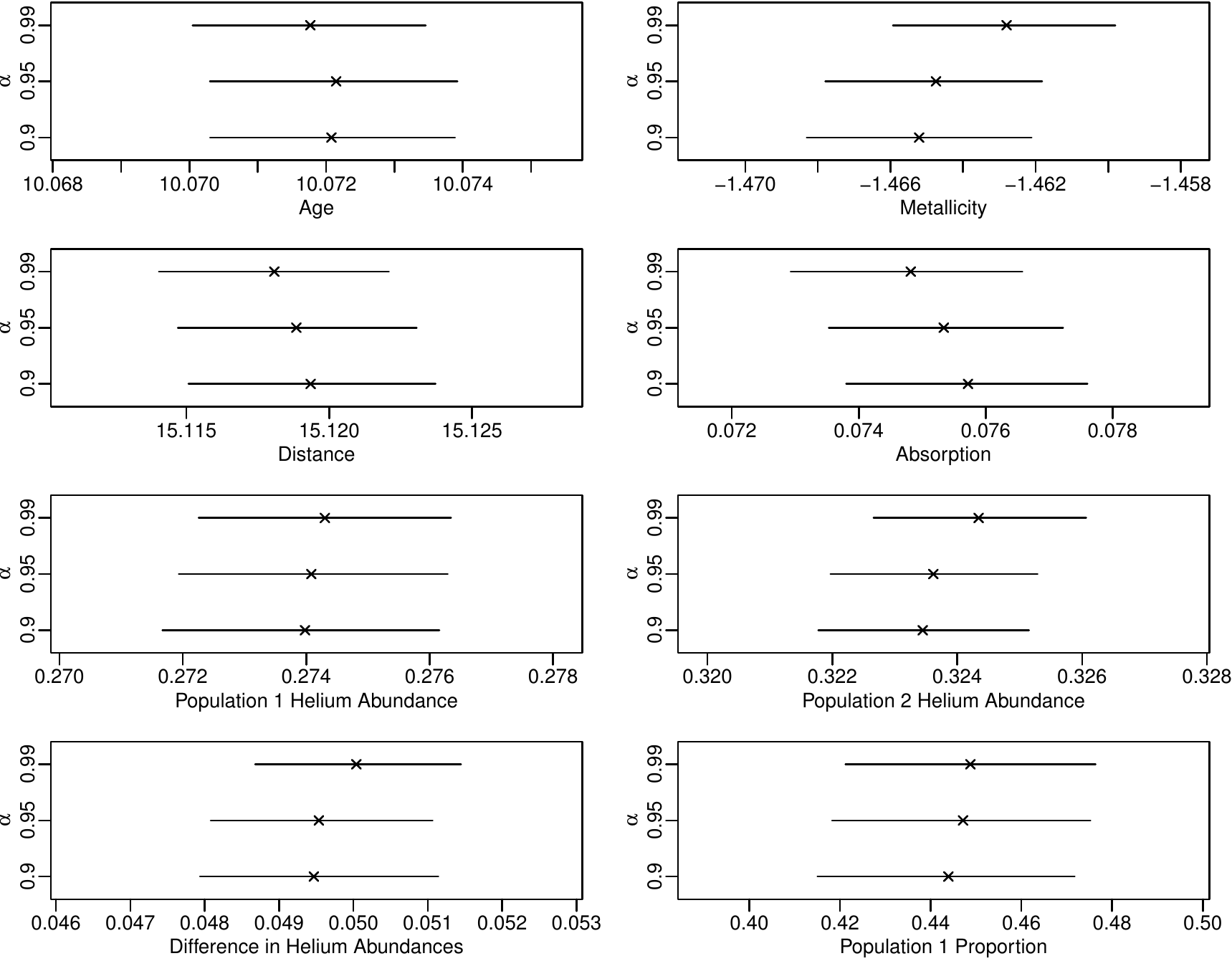}
\caption[Sensitivity analysis of $\alpha$ for NGC 5272] {Sensitivity analysis of $\alpha$ for NGC 5272.  Posterior means are denoted by an `x'.  The horizontal bars are 95\% posterior intervals.}
\label{fig:NGC5272_sensitivity}
\end{figure*}

For model fitting we assume all stars are singletons, which saves significant computation time and should offer a reasonable approximation because the expected percentage of binaries is only about 5\% \citep{Milone_2012b}.  The prior distributions for $\thetaFeH$, $\thetamod$, and $\thetaabs$ we use are 
\begin{align*}
	\thetaFeH &\sim N(-1.5, 0.05^2), \\
	\thetamod &\sim N(15.07, 0.05^2), \text{ and} \\
	\thetaabs &\sim TN(0.031, 0.01^2; 0),
\end{align*}
where $N(\mu, \sigma^2)$ is a Gaussian (i.e., Normal) distribution with mean $\mu$ and standard deviation $\sigma$, and $TN(\mu, \sigma^2; 0)$ is a Gaussian distribution with mean $\mu$ and standard deviation $\sigma$, truncated to be positive.  Prior means for $\thetaFeH$, $\thetamod$, and $\thetaabs$ come from the updated \citet{Harris_1996} globular cluster catalogue, with standard deviations chosen to be relatively conservative.  Ancillary information such as proper motions will eventually allow us to specify $P(Z_i=1)=\alpha_{i}$ on a star-by-star basis.  For now, however, we set $P(Z_i=1)=\alpha$ for all $i=1,\dots, N$ and investigate the sensitivity of results to $\alpha$.  Because we do not expect the fraction of field stars to be lower then 1\% or higher than 10\% we repeat our analysis with $\alpha$= 0.9, 0.95, and 0.99.  To fit each of the three resulting models, we run our AM algorithm for 30,000 iterations after the tuning period.  Inspection of the trace plots shows that every chain converges to its apparent stationary distribution by iteration 5,000.  We discard the first 5,000 iterations as burn-in, and base inference on the remaining 25,000 MCMC draws.  The results of the sensitivity analysis are presented in Figure~\ref{fig:NGC5272_sensitivity}; posterior means are indicated by an `x', and the horizontal bars are 95\% posterior intervals.  While the choice of $\alpha$ has a noticeable effect on the results, the effect is small and not scientifically meaningful.  We therefore use $\alpha=0.95$ for the remainder of our analysis.   

\begin{figure*}[t!]
\includegraphics[width=0.95\textwidth]{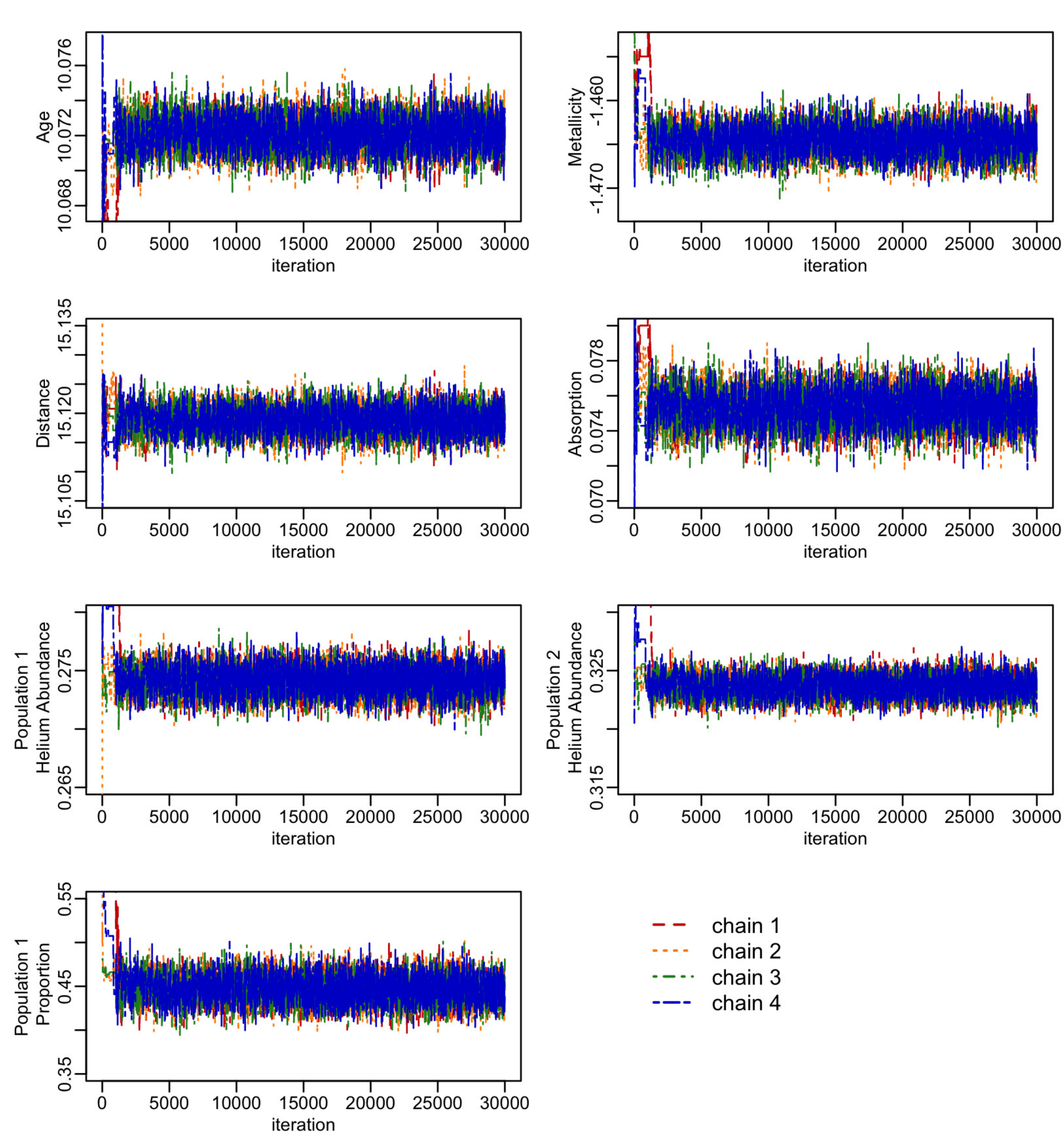}
\caption{The four chains we use to explore $P(\bTheta, \bPhi | \bX)$.  All chains reach their apparent stationary distribution well before iteration 5,000.}
\label{fig:NGC5272_chains}
\end{figure*}

After specifying $\alpha$, we explore $P(\bTheta, \bPhi | \bX)$ using four separate chains with different starting values.  This is done to diagnose proper convergence; if all chains eventually converge to the same distribution then our results are robust both to the starting values and to Monte Carlo variability among the chains.  Each chain is run for 30,000 iterations after the tuning period.  Inspection of the trace plots shows that every chain converges to the same apparent stationary distribution by iteration 5,000; see Figure~\ref{fig:NGC5272_chains}. For each chain we discard the first 5,000 iterations as burn-in, and keep the remaining 25,000 iterations. \changes{We also compute the Gelman-Rubin diagnostic \citep{Gelman_1992} on the post-burn-in iterations for each parameter, and all $\hat{R}$ values are equal to 1.\footnote{We use the {\tt gelman.diag} function (with {\tt autoburnin=FALSE}) in the {\tt coda} package from the {\tt R} programming language to compute the Gelman-Rubin diagnostic, $\hat{R}$, also known as the ``potential scale reduction factor." Values of $\hat{R}$ substantially above 1, e.g. greater than 1.1, indicate a lack of convergence.}} Fitted values and 95\% intervals for $(\bTheta, \bPhi)$, as well as for the difference in helium abundance, $\phi_{Y2} - \phi_{Y1}$, are given in Table~\ref{tbl:NGC5272_results}.  The fitted values are posterior means based on the 100,000 MCMC draws pooled from all four chains.  The reported 95\% credible intervals are the 2.5\% and 97.5\% posterior quantiles of these draws.  

\begin{table}[t!]
\begin{center}
\caption{Parameter Estimates for NGC 5272
\label{tbl:NGC5272_results}}
\begin{tabular}{lcc}
\hline
	Quantity & Fitted Value & 95\% CI \\ 
\hline \\
	$\thetaage$ & 10.072 & (10.070, 10.074) \vspace{0.1cm} \\ 
	$\thetaFeH$ & -1.465 & (-1.468, -1.462) \vspace{0.1cm} \\ 
	$\thetamod$ & 15.119 & (15.115, 15.123) \vspace{0.1cm}  \\
	$\thetaabs$  & 0.075 & (0.073, 0.077) \vspace{0.1cm} \\ 
	$\phi_{Y1}$ & 0.274 & (0.272, 0.276) \vspace{0.1cm} \\ 
	$\phi_{Y2}$ & 0.324 & (0.322, 0.325)  \vspace{0.1cm} \\ 
	$\phi_{Y2} - \phi_{Y1}$ & 0.0495 & (0.0481, 0.0511) \vspace{0.1cm} \\ 
	$\phi_{p1}$ & 0.447 & (0.419, 0.475) \vspace{0.1cm} \\ 
\hline
\end{tabular}
\end{center}
\vspace{-0.25cm}
\end{table}

\begin{figure*}[hp!]
\begin{center}
\includegraphics[width=\textwidth]{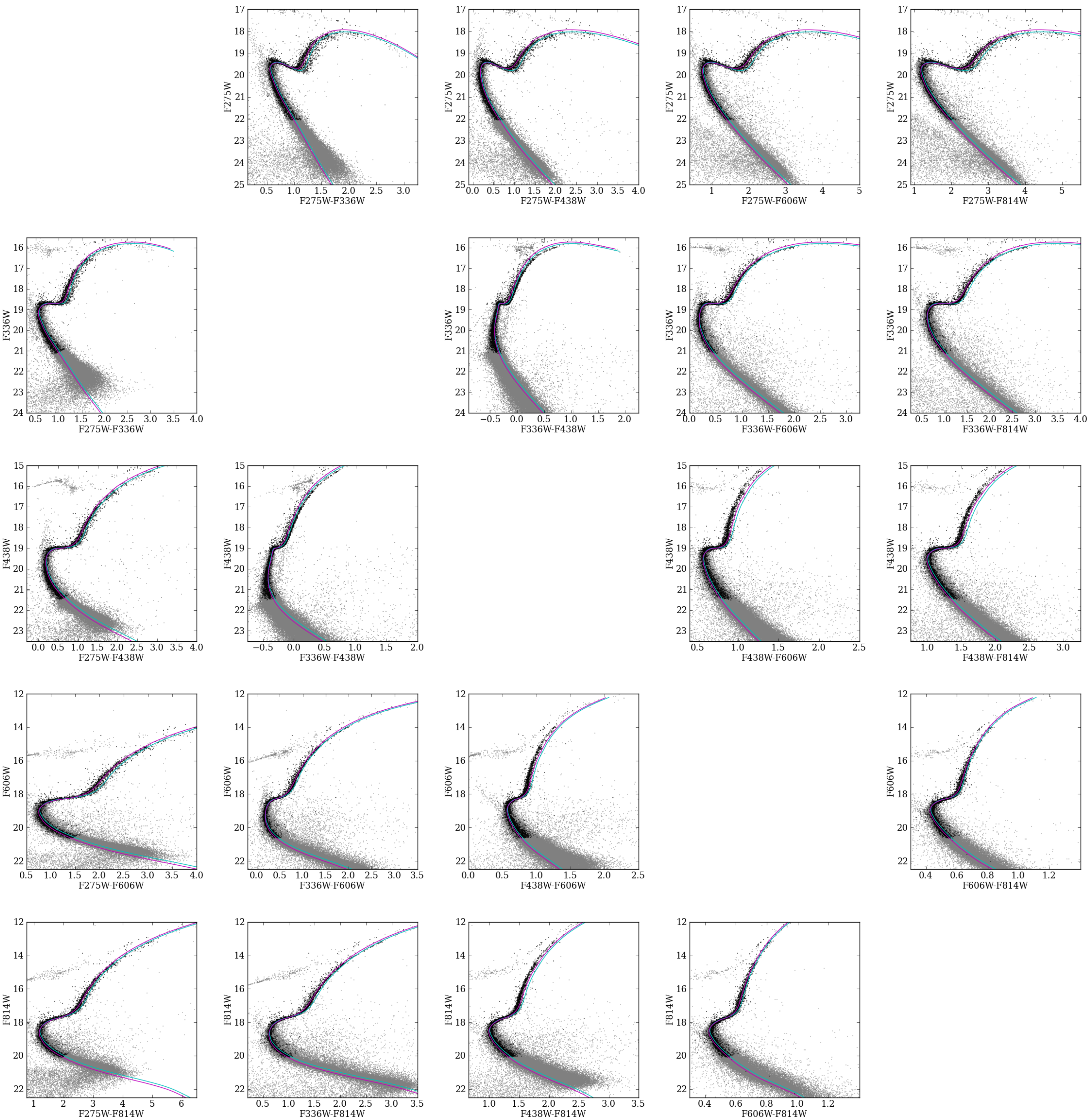}
\end{center}
\caption[Fitted model with all CMDs for NGC 5272] {Fitted model with CMDs for NGC 5272 constructed with all pairs of photometric magnitude bands.  Stars used in our analysis are represented in black; those removed are in grey.  Fitted isochrones for Populations 1 and 2 are represented by cyan and purple curves, respectively.}
\label{fig:NGC5272_fittedCMD}
\end{figure*}

In Figure~\ref{fig:NGC5272_fittedCMD} we present a matrix with CMDs for NGC 5272 constructed with all pairs of photometric magnitude bands, along with the fitted isochrones.  The fitted isochrone for Populations 1 and 2 are represented by cyan and purple curves, respectively.  It is clear that the fitted isochrones match the observed data well in some CMDs, and poorly in others.  In particular, CMDs that incorporate $F438W$ do not tend to be well fit.  This suggests subtle inconsistencies in the stellar evolution model that depend on wavelength; we discuss this further in Section~\ref{sec:MultiPopDiscussion}.  Examining these inconsistencies is a useful first step towards designing computer models that can better predict the observed data.

\section{Summary and Discussion}
\label{sec:MultiPopDiscussion}

In this article we present a Bayesian approach for fitting two-population globular clusters.  This is a substantial improvement over the common approach of plotting computer-model predictions on top of observed data and tuning the parameters until the two appear to agree.  By formulating a Bayesian model, we do not need to rely on any or all two-dimensional projections of five-dimensional data during model fitting.  This is important for fitting multiple-population clusters because the populations overlap in complex and non-obvious ways in CMDs.  We demonstrate with a simulation study that our method can adequately recover the population parameters of two-population clusters.  Specifically, we successfully recover the proportion of stars in each population and the difference in helium abundance between populations.  We also demonstrate how to diagnose model misspecification in the event that our two-population model is applied to a single-population cluster.  In particular, we show that (i) the fitted value for the proportion of stars in Population 1 is close to zero or one, and/or (ii) the posterior interval for the proportion extends over most of the range from zero to one.  

After demonstrating the capabilities of our two-population model, we analyze NGC 5272 as a proof of concept.  We verify that the value we specify for $\alpha$ is not overly influential, and explore the marginal posterior distribution of the cluster and population parameters using an AM algorithm that we devised for this purpose; the AM algorithm greatly improves convergence compared to its precursor non-adaptive Metropolis algorithm.  To diagnose convergence we run four separate chains per cluster, all with different starting values.  We find that the four separate chains quickly converge to the same apparent stationary distribution.

In addition to estimating the difference in helium abundance in NGC 5272, a secondary objective is to examine the model fits and investigate properties of the underlying stellar evolution model.  In general, we find that the fitted models do not agree with the observed data in CMDs involving the $F438W$ filters.  This disagreement is perhaps not surprising as model development follows observations, and data are only recently available in some of these HST passbands.  While we cannot conclude solely on the basis of our analysis of NGC 5272 that mismatch between fitted models and observed data is due to systematic errors in the computer model, further examination is warranted.  If this pattern persists with additional clusters and with verified photometric errors it may be that the morphologies in the computer model differ systematically from those in observed data; such a discrepancy has been discussed for fainter main sequence stars \citep{vanDyk_2009, DeGennaro_2009}.  Like any model fitting technique, our Bayesian approach relies on the accuracy of the underlying stellar evolution models.  Nevertheless, imperfect results can provide key feedback for improving the underlying models. 

Having demonstrated the capabilities of our model and methods for two-population globular clusters, work will focus on deploying them on many additional clusters.  Subsequently, we will extend our technique to include more than two stellar populations per cluster and incorporate additional population-level parameters such as the carbon, nitrogen, and oxygen abundances.  It is only by pairing such principled statistical approaches with recent high-quality HST visual/UV observations that we can estimate and interpret the parameters of multiple-population globular clusters.
\hfill \break
\hfill \break

This material is based upon work supported by the National Aeronautics \& Space Administration under Grant NNX11AF34G issued through the Office of Space Science. In addition, this project was supported by the National Aeronautics \& Space Administration through the University of Central Florida's NASA Florida Space Grant Consortium. DS was supported by NSF grant DMS 1208791 and the European Research Council via an Advanced Grant under grant agreement no. 321323-NEOGAL. DvD was partially supported by a Wolfson Research Merit Award (WM110023) provided by the British Royal Society and by Marie-Curie Career Integration (FP7-PEOPLE-2012-CIG-321865) and Marie-Skodowska-Curie RISE (H2020-MSCA-RISE-2015-691164) Grants both provided by the European Commission.


\newpage
\bibliographystyle{apj}
\bibliography{references}
\clearpage

\appendix

\section{Adaptive Metropolis Tuning Period}
\label{a:tuning}

\begin{table}[t!]
\begin{center}
\caption{Tuning Period Scaling Factors
\label{tbl:ScalingFactors}}
\begin{tabular}{lr}
\hline
	range of $a$ & $\varsigma(a)$ \\ 
\hline \\
	$a > 0.9$ & $2$ \\ 
	$ 0.7 < a \leq 0.9$ & $1.8$ \\
	$ 0.5 < a \leq 0.7$ & $1.5$ \\
	$ 0.4 < a \leq 0.5$ & $1.2$ \\
	$ 0.15 \geq a < 0.2$ & $1/1.5$ \\
	$ 0.05 \geq a < 0.15$ & $1/1.8$ \\
	$a < 0.05$ & $1/2$ \\ 
\hline
\end{tabular}
\end{center}
\end{table} 

The tuning period for our AM algorithm proceeds as follows:

\begin{enumerate}

\item Set $j=0$. Draw $\bY^{(d+1)} \sim N(\bY^{(d)}, 25\bDelta^{(j)})$ for $d=1,\dots, 99$, where $\bY^{(1)}$ is the starting value of the chain and $\bDelta^{(0)}$ is a diagonal covariance matrix with fixed variances, both of which are specified by the user.  The constant factor of 25 is chosen so that the chain takes ``big steps'' to explore the parameter space.
 
\item Set $j=j+1$.  If $j=20$, go to Step 5.  Else, draw $\bY^{(100j+k)} \sim N(\bY^{(100j + k -1)}, 5\bDelta^{(j-1)})$ for $k=1, \dots 50$.  During these iterations the chain takes ``medium steps'' to explore the parameter space, which may assist in jumping between modes.  Next, draw $\bY^{(100j+k)} \sim N(\bY^{(100j + k - 1)}, \bDelta^{(j-1)})$ for $k = 51, \dots, 100$.

\item Calculate the acceptance rate, $a$, of iterations $100j+51$ to $200j$.  If $0.2 < a < 0.4$, proceed to Step 4.  Else, set $\bDelta^{(j)} = \varsigma(a)\bDelta^{(j-1)}$, where $\varsigma(a)$ is given in Table~\ref{tbl:ScalingFactors}, and return to Step 2. 

\item Set $\bDelta^{(j)} = \bDelta^{(j-1)}$, then set  $j = j+1$. Draw $\bY^{(100j+k)} \sim N(\bY^{(100j + k -1)}, \bDelta^{(j-1)})$ for $k=1, \dots, 100$ and calculate $a$ for iterations $100j+1$ to $200j$.  If $0.2 < a < 0.4$, proceed to Step 5.  Else, set $\bDelta^{(j)} = \varsigma(a)\bDelta^{(j-1)}$ and return to Step 2.

\item Discard the first 100 draws produced during Step 1 and calculate the empirical covariance matrix of all remaining draws, which is then denoted by $\bxi^{(1)}$.  Then terminate the tuning period.

\end{enumerate}

Once we have calculated $\bxi^{(1)}$ from the tuning period, the AM algorithm proceeds as described in Section~\ref{sec:adapMCMC}.

\end{document}